\documentclass[twocolumn]{emulateapj}

\usepackage{amsmath}
\usepackage{bbding}
\usepackage[dvips]{color}
\usepackage[colorlinks=true,linkcolor=blue,citecolor=blue]{hyperref}

\usepackage{graphicx}
\usepackage{lipsum}

\newcommand{\diffconst}{0.25 $\pm$ 0.03\%} 

\shorttitle{KIC 1255b Transmission Spectrum}
\shortauthors{Schlawin, Herter, Zhao, Teske, Chen}

\begin{document}


\title{Reduced Activity And Large Particles From the Disintegrating Planet Candidate KIC 12557548b}


\author{Schlawin, E.\altaffilmark{1,2}, Herter, T.\altaffilmark{1}, Zhao, M.\altaffilmark{3,4}, Teske, J. K.\altaffilmark{5,6}, Chen, H.\altaffilmark{7}}
\altaffiltext{1}{Astronomy Department, Cornell University, Ithaca NY 14853}
\altaffiltext{2}{Steward Observatory, University of Arizona, Tucson AZ 85719}
\altaffiltext{3}{Astronomy \& Astrophysics, Pennsylvania State University, University Park PA 16802}
\altaffiltext{4}{Center for Exoplanets and Habitable Worlds, University Park PA 16802}
\altaffiltext{5}{Carnegie DTM, Washington, DC 20015}
\altaffiltext{6}{Carnegie Origins Fellow, jointly appointed by Carnegie DTM \& Carnegie Observatories}
\altaffiltext{7}{Astronomy Department, Boston University, Boston, MA 02215}



\bibliographystyle{apj}

\begin{abstract}
The intriguing exoplanet candidate KIC 12557548b is believed to have a comet-like tail of dusty debris trailing a small rocky planet. The tail of debris scatters up to 1.3\% of the stellar light in the Kepler observatory's bandpass (0.42 $\mu$m to 0.9 $\mu$m). Observing the tail's transit depth at multiple wavelengths can reveal the composition and particle size of the debris, constraining the makeup and lifetime of the sub-Mercury planet. Early dust particle size predictions from the scattering of the comet-like tail pointed towards a dust size of $\sim$0.1$\mu$m for silicate compositions. These small particles would produce a much deeper optical transit depth than near-infrared transit depth. We measure a transmission spectrum for KIC 12557548b using the SpeX spectrograph (covering 0.8 $\mu$m to 2.4 $\mu$m) simultaneously with the MORIS imager taking $r'$ (0.63 $\mu$m) photometry on the Infrared Telescope Facility for eight nights and one night in $H$ band (1.63 $\mu$m) using the Wide-Field IR Camera at the Palomar 200-inch telescope. The infrared spectra are plagued by systematic errors, but we argue that sufficient precision is obtained when using differential spectroscopic calibration when combining multiple nights. The average differential transmission spectrum is flat, supporting findings that KIC 12557548b's debris is likely composed of larger particles $\gtrsim$ 0.5$\mu$m for pyroxene and olivine and $\gtrsim$ 0.2$\mu$m for iron and corundum. The $r'$ photometric transit depths are all below the average Kepler value, suggesting that the observations occurred during a weak period or that the mechanisms producing optical broadband transit depths are suppressed.
\end{abstract}

%
\maketitle

\section{Introduction}
Analysis of public data from the Kepler mission \citep{borucki2010} uncovered the exotic KIC 12557548 system, which contains a disintegrating rocky planet candidate ($\lesssim0.1 M\earth$) with an orbital period of 15.7 hours \citep{rappaport}. The planet's escaping debris cause a variable ($\lesssim$0.2\% to 1.3\%) broadband optical absorption with a constant period, unlike any of the thousands of previously discovered transiting exoplanets. The large optical transit depths of the disintegrating debris are in contrast to escaping winds from hot Jupiters, which are transparent in broadband optical light and can only be detected at narrow gaseous absorption lines \citep[e.g.][]{vidalmadjar}. The broadband spectral nature of the atmospheric absorption and scattering indicates that the material is composed of solid particles. The long duration of the transit compared to a planetary crossing time indicates that it has a tail of debris longer than the diameter of its host star. Furthermore, the discoveries of KOI 2700b \citep{rappaport2014KOI2700}, K2-22b \citep{sanchis-ojedak2-22}, KIC 8462852 \citep{boyajian846} and WD 1145+017 \citep{vanderburg2015wdDisintegrating} reveal that there are multiple planets with tails of dusty effluents and more may be discovered around nearby stars. These disintegrating planets provide an exciting promise for characterizing the cores of planets that are inaccessible in the solar system and complement the studies of planet composition using the pollution of white dwarf atmospheres, which characterizes the end state of planetary accretion of the bulk of an entire planet onto evolved stars \citep[e.g.][]{jura2003wdPollution}.

The destruction mechanism proposed by \citet{rappaport} and \citet{perez-becker} to explain the tail of solid debris is that a hydrodynamic wind of a metal-rich vapor condenses into grains after adiabatic cooling. The calculated mass loss rate suggests that KIC 12557548b may be in the catastrophic end state of its life, which will destroy it entirely \citep{perez-becker}. Stellar activity may factor into the planet's destruction -- the deepest transit depths during the Kepler observations tend to coincide with times at which the star spots face our line of sight and shallow transit depths occur when the stellar spots are rotated out of our line of sight \citep{kawahara2013starspots}. This correlation has been confirmed with additional Kepler data, but there is also the possibility that occultations of star spots by the comet's tail affect the transit depth \citep{croll2015starspots}. The transit depth traces the disintegrating activity directly as the planet's physical radius is too small to contribute more than $10^{-5}$\ to the average $10^{-2}$ transit depth.

The average Kepler light curve of KIC 12557548 over 2000 transits is a diagnostic tool for measuring the properties and grain sizes of KIC 12557548b's debris. Prior to the flux decrement (transit), there is a small flux increase due to scattering of dust particles into the line of sight. The flux increase is sensitive to dust particle size, and models matching the flux increase are consistent with Mie-scattering from small silicate dust grains ($\sim$ 0.1 $\mu$m) \citep{budaj12,brogi2012}. \citet{budaj12} modeled the light curve as Mie-scattering and absorption from spherical pyroxene and iron dust grains 0.01$\mu$m to 1$\mu$m in radius. The models with 0.1 $\mu$m to 1.0 $\mu$m grains fit the beginning of the average Kepler light curve better whereas the models with 0.01 $\mu$m to 0.1 $\mu$m grains fit the egress better, suggesting that larger particles either are ejected or form near the planet and get whittled down as they progress along the tail. \citet{brogi2012} model the scattering of dust particles with a Henyey-Greenstein phase function and find similar size particles, with a best-fit power law grain size distribution from 0.04$\mu$m to 0.19$\mu$m. KIC 12557548b's tail length also constrains the particle size and dust composition because the tail length is a strong function of the sublimation time of dust particles. \citet{vanlieshout2014kic1255comp} estimate the sublimation timescales for a variety of dust compositions and sizes and find that only 10$\mu$m Corundum and 100$\mu$m or larger silicates can have the right sublimation times at KIC 12557548b's orbital radius to reproduce the observed tail length.

A multi-wavelength simultaneous light curve provides additional diagnostics of the planet's escaping transiting debris because the scattering and absorption of dust particles is highly wavelength dependent when the wavelengths become larger than the size of the particles \citep[e.g.][]{draine11}. \citet{croll2014} obtained simultaneous Kepler ($\sim$0.65$\mu$m) and $K_S$-band (2.15$\mu$m) light curves for two separate epochs and compared the transit depths. Both the 0.65$\mu$m and 2.15$\mu$m light curves were consistent within errors, giving evidence for particle sizes $\gtrsim 0.5 \mu$m.

The size of particles escaping a disintegrating planet can change with time, so studies at different epochs may give different results. \citet{bochinski2015evolving} find that the spectral slope of the debris cloud changes between two nights of observation. The spectral slope between the $u'$, $g'$ and $z'$ bands of a deep transit is consistent with interstellar medium reddening laws for (0.25$\mu$m to 1$\mu$m) particles, whereas the shallow transit had a flatter spectral energy distribution, which would be expected for larger grains. Similarly, spectra of three different transits of K2-22b \citep{sanchis-ojedak2-22} show that the spectral slope during the deep transit ($\sim 0.8\%$) is consistent with dust particles $\sim 0.2 \mu$m to $\sim 0.4 \mu$m in size, and the two shallow transit events ($\sim 0.4\%$) are spectrally flat and consistent with larger-sized dust particles.

We observed the disintegrating system KIC 12557548b over 8 transits with the SpeX spectrograph and MORIS imager on the Infrared Telescope Facility (IRTF), covering the wavelengths from 0.6 $\mu$m to 2.4 $\mu$m, to give additional constraints on the particle sizes escaping from KIC 12557548b. Section \ref{sec:obs} describes our observations, where we show that the optical transit depths are all shallower than the average Kepler value and that the infrared spectrum is flat. Section \ref{sec:transm} describes our fitted transmission spectrum, which is best fit by large particles, $\gtrsim 0.5 \mu$m for silicate compositions and  $\gtrsim 0.2 \mu$m for iron and corundum compositions. We conclude in Section \ref{sec:kic1255conclusion}, and also consider different ways to average the variable spectra in Section \ref{sec:altAvgMeth}.

\section{Observations}\label{sec:obs}

\subsection{Observational Setup}

\begin{figure}[!ht]
\begin{center}
\includegraphics[width=0.39\textwidth]{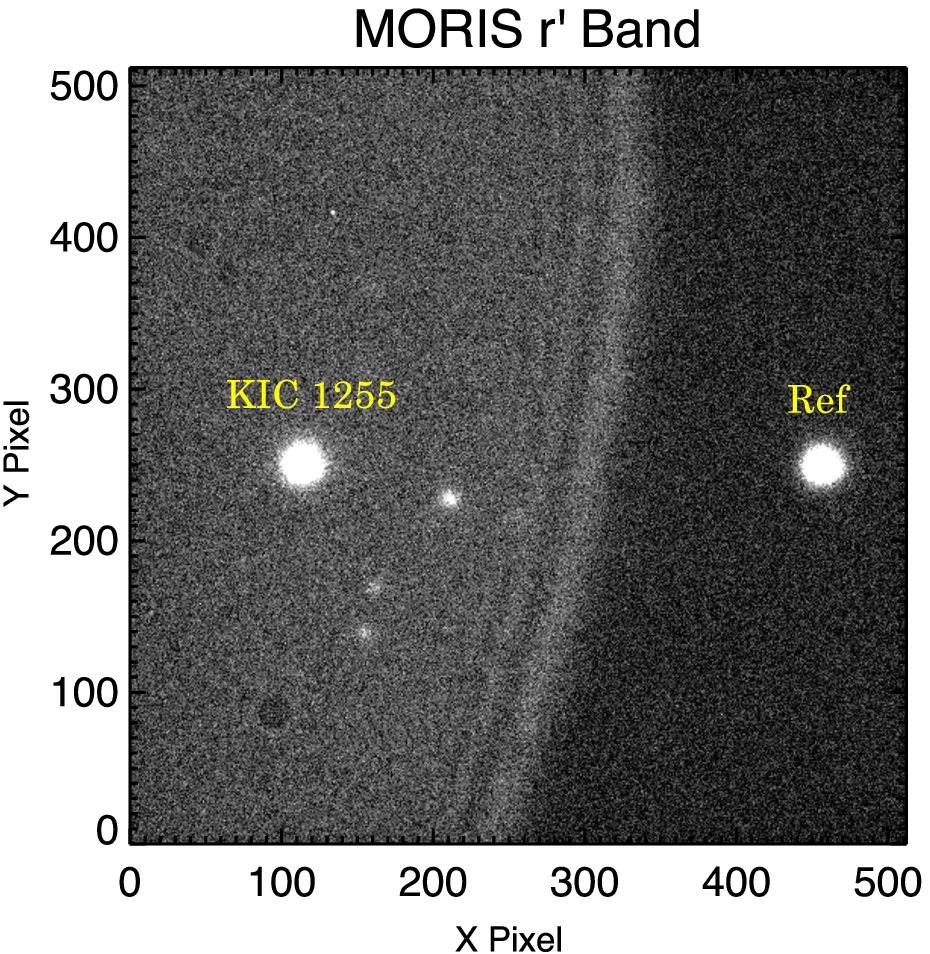}
\vspace{0.15in}
\\
\includegraphics[width=0.45\textwidth]{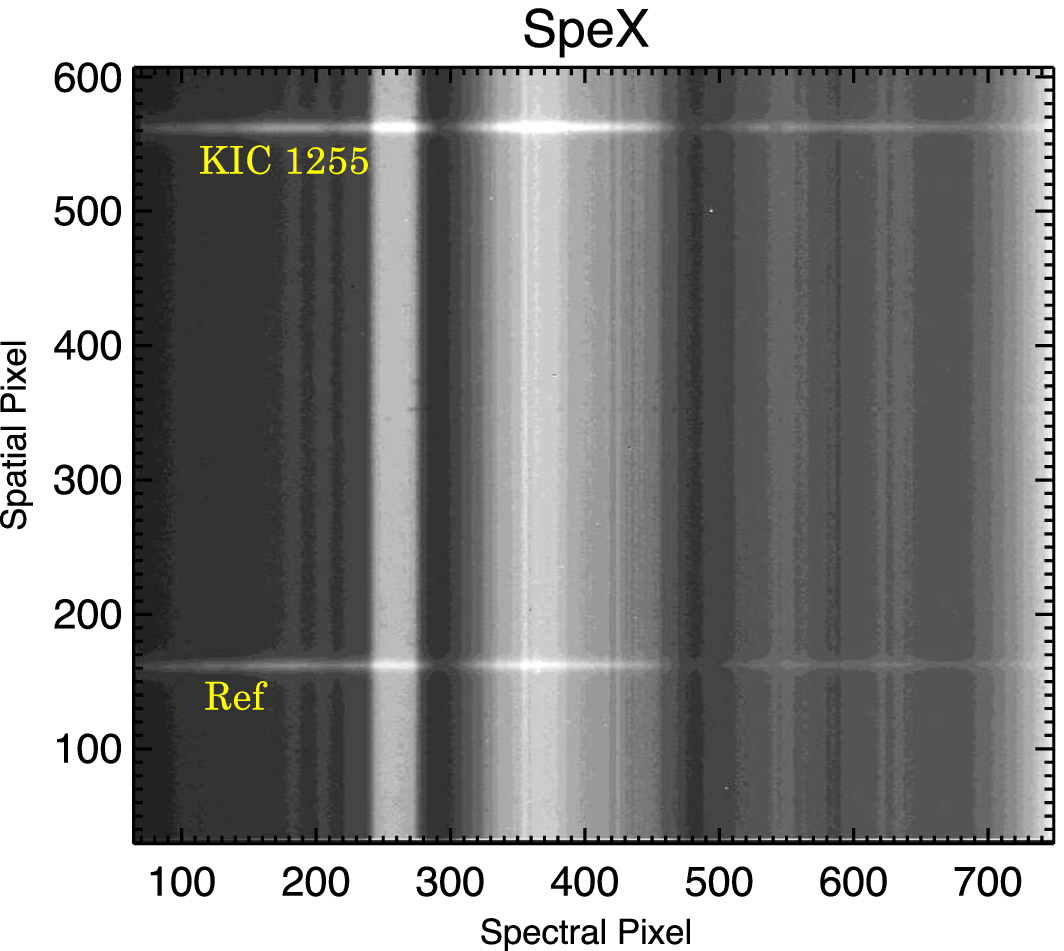}
\end{center}
\caption{{\it Top} Example raw MORIS image, showing an arc of emission that is likely a reflection off the foil that reduces the emissivity of the IRTF telescope. The contrast of the arc emission has been emphasized in this image for visual purposes although it is a small fraction of the stellar flux. {\it Bottom} Example SpeX image after rectifying all wavelengths for background subtraction.}\label{fig:exampleSpecMorisImg}
\end{figure}

\subsubsection{Reference Star Setup}

We observed 8 transits of KIC 12557548b with a simultaneous reference star (2MASS J19234770+5130175) to correct for telluric, detector and instrument systematics. This is the same general observational setup used when obtaining a transmission spectrum of the hot Jupiter CoRoT-1b \citep{schlawin2014}, but there are a few noteworthy differences. The Kepler observatory \citep[0.423 $\mu$m to 0.897 $\mu$m bandpass;][]{koch2010keplerChar} halted photometry after a reaction wheel failure, so for KIC 12557548 we used the MORIS imager \citep{Gulbis2011} in the $r'$ band (0.542 $\mu$m to 0.693 $\mu$m) to give optical information simultaneously with the SpeX infrared wavelengths 0.8$\mu$m to 2.4$\mu$m, shown in Figure \ref{fig:exampleSpecMorisImg}. KIC 12557548 ($K$ =13.32) and its reference star J19234770+5130175 \citep[K=14.00, as determined by 2MASS][]{skrutskie06}, are both fainter than CoRoT-1 (K=12.15) and its nearby reference star (K=11.50) meaning that the minimum photon-limited noise floor is higher and that background subtraction plays a larger role in the spectral extraction. The KIC 12557548 system is overall more challenging to achieve high precision on due to its faintness and the fact that the transit depths are variable, so multiple nights must be acquired to ensure transit detection.

The simultaneous MORIS imager $r'$ band photometry is achieved with a 0.8 $\mu$m dichroic to split off short wavelengths while still passing the infrared light to the spectrograph. The field of view of MORIS is 1' x 1', similar to that of the guide camera of SpeX and it permitted us to include the same reference star (2MASS J19234770+5130175)  on the MORIS detector. We used exposures of 5s in 2013 and later increased to 10 - 20s in 2014 to reduce seeing and scintillation noise (Table 1). The camera was kept in focus throughout the observations \citep[as opposed to a purposeful de-focusing technique, eg.,][]{southworth2009defocusing} due to KIC 12557548's faintness and the fine pixel sampling of MORIS (0.12\arcsec/pixel).

\subsubsection{Spatial and Temporal Baselines}
The reference star, 2MASS J19234770+5130175, shown in Figure \ref{fig:exampleSpecMorisImg}, is 39\arcsec\ away from KIC 12557548, so that both can lie in the 60\arcsec\ long slit with more than 5\arcsec\ of baseline for background subtraction. The upper baseline (in Figure \ref{fig:exampleSpecMorisImg}) is shorter for KIC 12557548 (5\arcsec) than for the reference star (16\arcsec) because the stars must be displaced upward on the SpeX array in order to fit both on the MORIS detector. In addition to this spatial baseline for background subtraction, the light curves require time baselines on either side of transit to fit the normalized out-of-transit flux. We obtained 0.5 to 2.6 hours of baseline on either side with an average of 1.4 hours, comparable to the $\sim$1.4 hour transit duration where the average Kepler light curve drops below 99.95\% in normalized flux. The seeing varied between all the nights, as shown in Table \ref{tab:obsSummary}, but was always smaller than 1/3 of the 3 arcsecond slit width. The widest available slit on IRTF is 3\arcsec. Furthermore, care was taken to align the two stars within the slit so that they were less than 0.1 arcsec displaced in the dispersion direction to minimize differential slit loss between KIC 12557548 and its reference star.

\begin{figure*}[!ht]
\begin{center}
\includegraphics[width=1.0\textwidth]{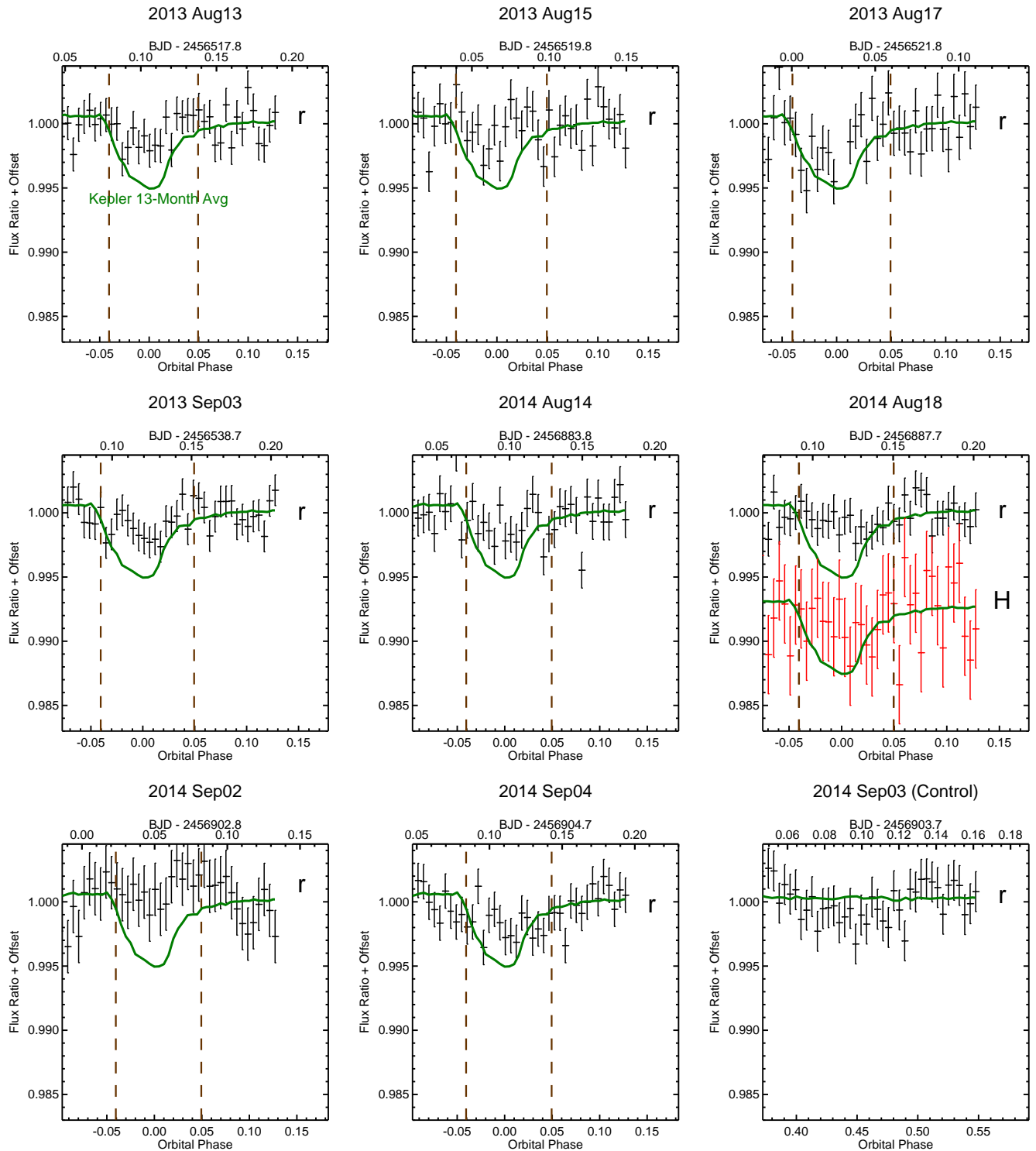}
\end{center}
\caption{The time series in the $r'$ (0.63$\mu$m) filter of the MORIS camera (black points with error bars) and $H$ band filter of the WIRC camera show transit depths that are shallower than the average short cadence Kepler light curve (green). The vertical dashed lines mark the approximate start and end of the transit (using the points where the average Kepler light curves falls below 99.95\%).}\label{fig:PhotIndNights}
\end{figure*}

The 8 transits are spread over many nights, spatial positions, detectors and potential disintegration activity of the planet. The four observations in 2013 on UT August 13, 15, 17 and UT September 3 were followed by a SpeX detector upgrade from a Aladdin InSb detector to a Hawaii-2RG with lower read noise. The new Hawaii-2RG is sensitive to alpha particles emitted by a Thorium-containing coating within the SpeX instrument but they are easily excluded with 5$\sigma$ clipping in the spatial profile fitting and time series. All 2013 observations were interrupted by a telescope guiding jump subsequently determined to be due to a chopping mirror; these jumps were avoided in 2014. The interruptions in 2013 were corrected for within 10 minutes, with the exception of UT August 17, which also suffered a telescope guiding issue that had to be corrected at zenith, visible in Figure \ref{fig:PhotIndNights}. In 2014, we observed the transit on UT August 14, August 18, September 2 and September 4 with no guiding interruptions, good seeing and lower read noise in the newer Hawaii-2RG detector. We also observed the same system on UT September 3, 2014 when it was outside of transit to quantify the light curve systematics over 3 hours and to use as a control.

\begin{table}
	\begin{center}
	\begin{tabular}{rrrrcc}
	UT Date	     & $t_\mathrm{spec}$	 & $ t_\mathrm{phot}$ & Seeing & Airmass & Sun Alt \\
	 		     & (s)				&	(s)			&   (\arcsec) &  & ($\deg$) \\
	\hline \hline
	08/13/13 & 75			&  5				& 0.5 & 1.18 $\rightarrow$ 1.18 $\rightarrow$ 1.62 &  -47 $\rightarrow$ -49 \\
	08/15/13 & 75			&  5				& 1.0 & 1.19 $\rightarrow$ 1.17 $\rightarrow$ 1.40 &  -35 $\rightarrow$ -55 \\
	08/17/13 & 75			&  5				& 0.6 & 1.32 $\rightarrow$ 1.17 $\rightarrow$ 1.27 &  -15 $\rightarrow$ -56 \\
	09/03/13 & 75			&  5				& 0.4 & 1.21 $\rightarrow$ 1.17 $\rightarrow$ 1.45 &  -18 $\rightarrow$ -61 \\
	08/14/14 & 60			&  10				& 1.0 & 1.19 $\rightarrow$ 1.17 $\rightarrow$ 1.53 &  -34 $\rightarrow$ -52 \\
	08/18/14 & 60			&  20				& 0.8 & 1.27 $\rightarrow$ 1.17 $\rightarrow$ 1.46 &  -19 $\rightarrow$ -56 \\
	09/02/14 & 60			& 20				& 0.6 & 1.19 $\rightarrow$ 1.17 $\rightarrow$ 1.72 &  -24 $\rightarrow$ -61 \\
	09/04/14 & 60			& 20				& 0.5 & 1.23 $\rightarrow$ 1.17 $\rightarrow$ 1.74 &  -14 $\rightarrow$ -62 \\
	\hline
	\multicolumn{6}{c}{Out of Transit} \\
	09/03/14\* & 60			& 20				& 0.5 & 1.26 $\rightarrow$ 1.17 $\rightarrow$ 1.30 &  -10 $\rightarrow$  -56 \\	
	\end{tabular}
	\end{center}
	\caption{\rm Summary of the 8 transits and 1 out-of-transit nights, with the exposure time for SpeX spectra $t_\mathrm{spec}$, MORIS photometric exposure time $t_\mathrm{phot}$ and FWHM median seeing in arc seconds measured across the infrared wavelengths from 0.9$\mu$m to 2.3$\mu$m. The airmass column shows the starting, minimum and ending airmass. The sun altitude column shows the starting and ending sun altitudes.}\label{tab:obsSummary}
\end{table}

\subsection{Photometric Reduction}
\subsubsection{MORIS Photometry}

Photometric data reduction was carried out following the pipeline and steps of \citet{zhao12}. We found that an aperture size of 28 pixels in diameter (corresponding to 3.4\arcsec) and a 35-pixel wide background annulus provided the lowest out-of-transit scatter in the light curves for all 2013 nights, although aperture sizes with $\pm3$ pixels gave the same results. In 2014, the reference star was closer to the edge of the MORIS detector because we moved the stars to a a cleaner part of the 3\arcsec $\times$ 60\arcsec\ slit for SpeX. Thus, a different method was required to extract photometry. We subtracted KIC 12557548 by the reference star and extracted photometry on the difference image normalized by the photometry on the reference star. This allowed a similar level of precision as in 2013 but using a 30 pixel aperture diameter and only a 14 pixel wide background annulus, separated from the aperture by 4 pixels.  A non-uniform background illumination (a 1\% step in background brightness going from the reference star to target star with a peak at the intersection, visible in Figure \ref{fig:exampleSpecMorisImg}) plagued the MORIS camera on all nights. The likely cause of this non-uniform background is a specular reflection from a bright star. The structure in the background moves during the night but the step in brightness remains between the stars, so that the background is relatively uniform surrounding each star when performing aperture photometry. This background structure is not visible in the infrared slit viewer within SpeX nor the spectrograph, likely because it is mitigated by baffling or the cold pupil stop in SpeX.

\subsubsection{WIRC Photometry}

On UT August 18, 2014, we also simultaneously observed a transit of KIC 12557548b with the Palomar Hale 200-inch telescope and its Wide-field Infra Red Camera (WIRC) \citep{wilson2003wirc} in the $H$ band. We used an engineering-grade HAWAII-2 array with a field of view of $8.7' \times 8.7'$ for the observation\footnote{The original science grade H2 array failed in April 2014. The engineering grade array has much lower well depth, more hot/bad pixels, and one less-responsive quadrant, but is nonetheless sufficient for this study due to the faintness of the target and enough reference stars nearby.}. We followed the well-established precision photometry procedures for WIRC and ``stared" at the target throughout the observation \citep[e.g.,][]{zhao12, zhao2014hatp32}. We took 30s exposures and kept the telescope focused due to the faintness of the target and the high background.  For the same reason, we also took a sky background immediately before and after the observation to construct a normalized ``supersky" for sky subtraction in the reduction, which was necessary to remove large scale structures and detector fringing residuals not removed by a dome flat. We followed the data reduction steps outlined in \citet{zhao2014hatp32}, and used 6  stars with fluxes in the range of $0.6-2\times$ that of the target in the good part of the detector as references. A photometry aperture of 5\arcsec~ and a background annulus from 6.3\arcsec~ to 10\arcsec~ resulted in the lowest RMS residuals. We also checked that this photometric aperture does not introduce any systematics, such as can happen when simply minimizing the RMS \citep{croll2015emissionHJ}.

\subsubsection{Comparison with Kepler Photometry}
The optical transit depths, as measured by MORIS, are much smaller than the past behavior for the planet as determined by the Kepler observatory, as shown in Figure \ref{fig:PhotIndNights}. The $r'$ band transit depths for 8 nights are {\it all} less than 0.43\%, whereas the {\it average} Kepler value is 0.504\%. If we assume transits are all statistically independent and have a probability distribution function as given in \citet{vanWerkhoven2014}, the probability of 8 transits less than 0.43\% is 0.2\% (the equivalent of a 3.2 $\sigma$ event in a Gaussian distribution), suggesting that the activity fell into the shallow periods found by Kepler ($\lesssim 0.1\%$ depths) or that it changed after May 2013, when the Kepler observatory stopped monitoring KIC 12557548. KIC 12557548b was observed by the Kepler observatory to have $\sim$20 day intervals of weak activity with shallow transit depths ($\lesssim 0.2\%$) \citep{vanWerkhoven2014}. However, it would be unlikely that the observations in both 2013 and 2014 would fall into shallow transit intervals as there were only two 20 day weak activity intervals in 3.7 years of monitoring in the 15 quarter analysis by \citet{vanWerkhoven2014}. 

\subsection{Spectroscopic Reduction}
The processing pipeline for the SpeX spectrograph used in \citet{schlawin2014} was improved to handle the relatively larger background-to-source brightness. The spectrograph images (prior to extraction) were rectified so that each column corresponds to one wavelength (see Figure \ref{fig:exampleSpecMorisImg}), but instead of using an Argon lamp spectrum for the straightening, the background spectrum was cross-correlated with a master image to account for the slight differences in illumination and flexure between the Argon lamp and sky image. We dark-subtracted each frame and flat-fielded using a sky flat to fit the non-uniform transmission of the reflective 3\arcsec $\times$ 60\arcsec\ slit and a filtered lamp image to remove the pixel-to-pixel responsiveness variations. The sky flats were shifted to account for the flexure of the instrument. The image rectification and sky flat fields improved background subtraction while still using a simple 4th-order background fit including all pixels more than 3.7 arcseconds from the source on either side.

We employed a custom optimal extraction based on the method of \citet{horne1986optimalE}, to increase control of the spatial profile polynomial fit and we use the standard deviation of the background fit residuals for empirical estimation of background and read noise. We use a 7 arcsecond aperture, which is well above the  seeing for all nights (0.5 to 1 arcsec in the infrared, listed in Table \ref{tab:obsSummary}). We tested aperture sizes from 5 to 10\arcsec\ with little effect on the results because optimal extraction weights the core of the PSF more than the wings. This pipeline is checked on CoRoT-1b transit data \citep{schlawin2014} and produces consistent results, but the improvements are less important for the higher signal-to-background ratio of the CoRoT-1b observations.

\section{Differential Dynamic Spectrum}\label{sec:transm}

\begin{figure}[!ht]
\begin{center}
\includegraphics[width=0.5\textwidth]{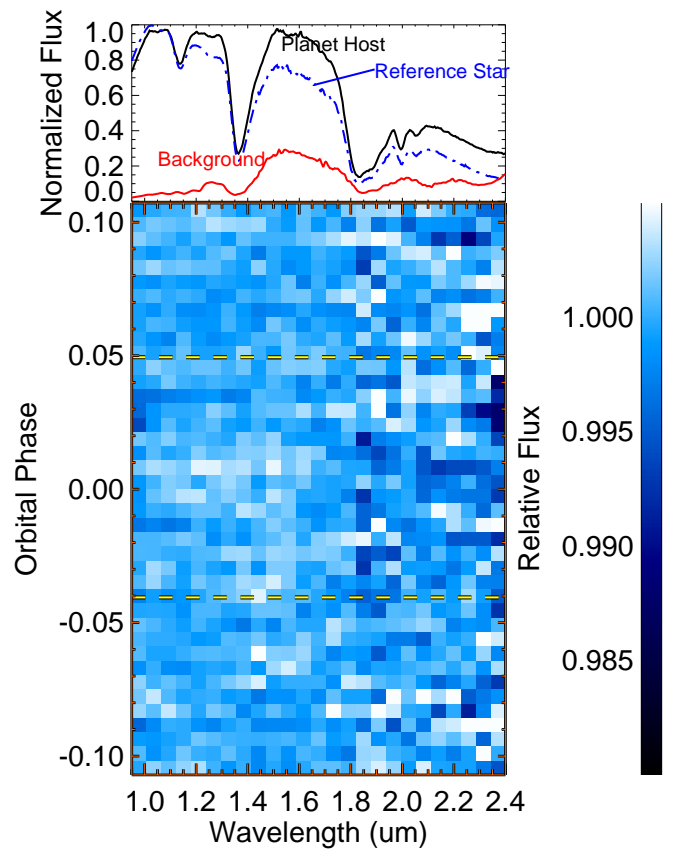}
\end{center}
\caption{{\it Top Panel} Normalized KIC 12557548, reference star and background spectra showing where the signal to noise drops because of telluric (Earth's) absorption and background emission. {\it Bottom Panel} Dynamic differential spectrum for the combined 6 nights of transit with 25 wavelength bins and 5$\sigma$ clipping to remove cosmic rays, alpha particle hits and large systematics. There is no statistically differential transit at a phase of 0.0.}\label{fig:dynamicSpecMultiN}
\end{figure}

There are substantial common-mode (affecting all wavelengths) systematics affecting the SpeX detector on most nights, including August 18, 2014 where we obtained simultaneous $H$ WIRC photometry. For this night, the $H$ band transit depth is 0.19 $\pm$ 0.11 \% consistent with the $r'$ depth of 0.07\% $\pm$ 0.06\% whereas the SpeX spectra binned to the same wavelengths as $H$ band (1.48 to 1.78 $\mu$m) give 1.26 $\pm$ 0.18\%. The rest of the wavelengths measured by the SpeX spectrograph are also 1.3\% to within errors, suggesting that it has a common-mode systematic. There are fluctuations in the broadband transit depth measured by the SpeX instrument, going from a large flux decrease during transit on August 15, 2013 to a flux increase during transit on August 17, 2013. We therefore analyze the SpeX data in a differential way by dividing all light curves by the binned broadband time series (differential spectra). The absolute spectra, alternate averaging methods and transit injection tests are explored in Appendix \ref{sec:altAvgMeth}.

To produce the differential spectra, we perform the following steps:
\begin{enumerate}
\item Divide the planet host star spectrum by the reference star spectrum to obtain a corrected spectrum for one single image
\item \begin{enumerate}
	\item Bin the corrected spectrum into equally spaced wavelength bins
	\item Average the separate wavelength bins to get a broadband flux point
	\item Divide each of wavelength bins by the broadband flux point to get a relative (differential flux) for that image
	\end{enumerate}
\item Repeat for all images to build up a separate time series for each wavelength bin
\end{enumerate}

The systematics present in the absolute SpeX time series that occurred for KIC 12557548 but not CoRoT-1b \citep{schlawin2014} were surprising, so we performed numerous tests on the data to ascertain the best averaging and analysis methods. Most spectroscopic exoplanet observations are affected by systematics like detector ramps \citep[e.g.][]{crossfield2012spitezerMIPS}, detector non-linearity \cite[e.g.][]{gibson12}, variable slit loss \citep[e.g.][]{sing12} and telluric absorption \citep[e.g][]{crossfield2012wasp12}, some of which can be calibrated out and others not. We searched the SpeX observations for evidence of differential slit loss, specks on the reflective slit substrate, non-linearity, detector anomalies, guiding errors, background variations, stray light, spectral shifting, insufficient OH airglow subtraction, sum versus optimal extraction, extraction aperture size, background fitting order, background fitting regions, flat field errors, telluric absorption lines, correlations with airmass and telescope flexure. None of these tests resulted in robust correlations with the observed anomalies.

One of the expected culprits for systematic effects on the light curve is slit loss between the reference star and target star. Indeed, when we use the wings of the PSF from the MORIS imager to estimate the slit loss of a 3\arcsec\ slit by extracting photometry outside of a 3\arcsec\ extraction box, we find that about 7\% absolute slit losses are possible in the optical. However, what matters for the time series precision is the relative slit losses between the target and reference star. In \citet{schlawin2014}, sub-0.1\% transit depth precision was achieved for the reference star 31\arcsec\ away, suggesting that differential slit losses do not contribute significantly to the large observed systematics. The night with the largest light curve variations, August 15, 2013, has calculated differential slit losses between the two stars ($< 1\%$) from the MORIS imager spatial profiles that are much less than the observed variations ($\sim$5\%). Furthermore, the systematic effects should decrease with longer wavelengths (where the seeing is better) if they were related to slit losses, whereas the systematics tend to be worst at the longest wavelengths. For these reasons, we conclude that slit loss is not a dominant cause of the observed systematics.

The control night of September 3, 2014 suggests that non-astrophysical variations are possible during transit. We fit the light curves on this control night in the same manner as Section \ref{sec:lcFit} as if it were a real transit by shifting the time by 0.3 days. We find that for the five wavelength channels, the transit depths deviate from zero -- see Table \ref{tab:controlSpec}. Although there is still a trend in this control night from the shortest to longest wavelengths, the differential transmission spectrum shows smaller deviations from a flat spectrum than the absolute transmission spectrum. We therefore argue the systematic errors are better mitigated by dividing all SpeX data by the average light curve to remove common-mode errors. Furthermore, combining data from multiple nights averages out the random fluctuations that can happen in a single time series, since each night has its own unique set of spatial position, sun angle, starting airmass, seeing and sky brightness. We therefore choose to fit differential spectra for each night and average them together to derive the planet system's transmission spectrum.

\begin{table}
	\begin{center}
	\begin{tabular}{l*{1}rr}
	Wavelength   & Absolute 				& Differential \\
	 ($\mu$m)	     & Transit Depth (\%)		& Transit Depth (\%) \\
	\hline \hline
	  0.63 ($r'$)   &	0.16		$\pm$  0.06 	& \\			
	  0.98  	    & 	-0.21		$\pm$  0.13	& -0.50 $\pm$ 0.18 \\		
	   1.29           & 	-0.09		$\pm$  0.23	& -0.29 $\pm$ 0.21 \\		
	   1.61 	    & 	-0.14		$\pm$  0.15	& -0.33 $\pm$ 0.12 \\		
	   1.93          & 	0.26		$\pm$  0.26	& 0.02 $\pm$ 0.24 \\			
	   2.24          & 	0.84		$\pm$  0.23	& 0.60 $\pm$ 0.25		
	\end{tabular}
	\end{center}
	\caption{\rm Fitted transit depths for the control night of September 3, 2014, showing systematic deviations from zero. The absolute transit depths show variations up to 3.7$\sigma$ from zero whereas the differential transit depths stay within 3$\sigma$ from average. We adopt the method of dividing all light curves by the broadband time series (differential time series) and combining multiple nights together when analyzing the SpeX data to remove the common-mode systematic errors.}\label{tab:controlSpec}
\end{table}

Figure \ref{fig:dynamicSpecMultiN} shows the differential dynamic spectrum surrounding mid-transit of KIC 12557548b where the 5 nights for which MORIS transit depths were greater than 0.1\% (August 13, 2013, August 17, 2013, September 03, 2013, August 14, 2014 and September 04, 2014) are combined into 25 wavelength bins and 5.4 minute time bins. Prior to combining the 5 nights together, the individual nights have broadband behavior divided out as described in the three step list above. The orbital phase is centered on a reference epoch of BJD=2454833.039, the flux minimum of the average Kepler light curve using the publicly available data. The approximate transit start and end (where the flux drops below 99.95\% of the out-of-transit flux as measured in the average Kepler transit profile) are shown as horizontal dashed lines. The different nights are averaged with 5$\sigma$ clipping from the median to remove bad pixels, cosmic ray hits and the alpha particle hits caused by a Thorium anti-reflective coating within SpeX. The differential dynamic spectrum shows no statistically significant deviations from a flat spectrum (ie. all points are within 1.1$\sigma$ from the mean).

\begin{figure}[!ht]
\begin{center}
\includegraphics[width=0.48\textwidth]{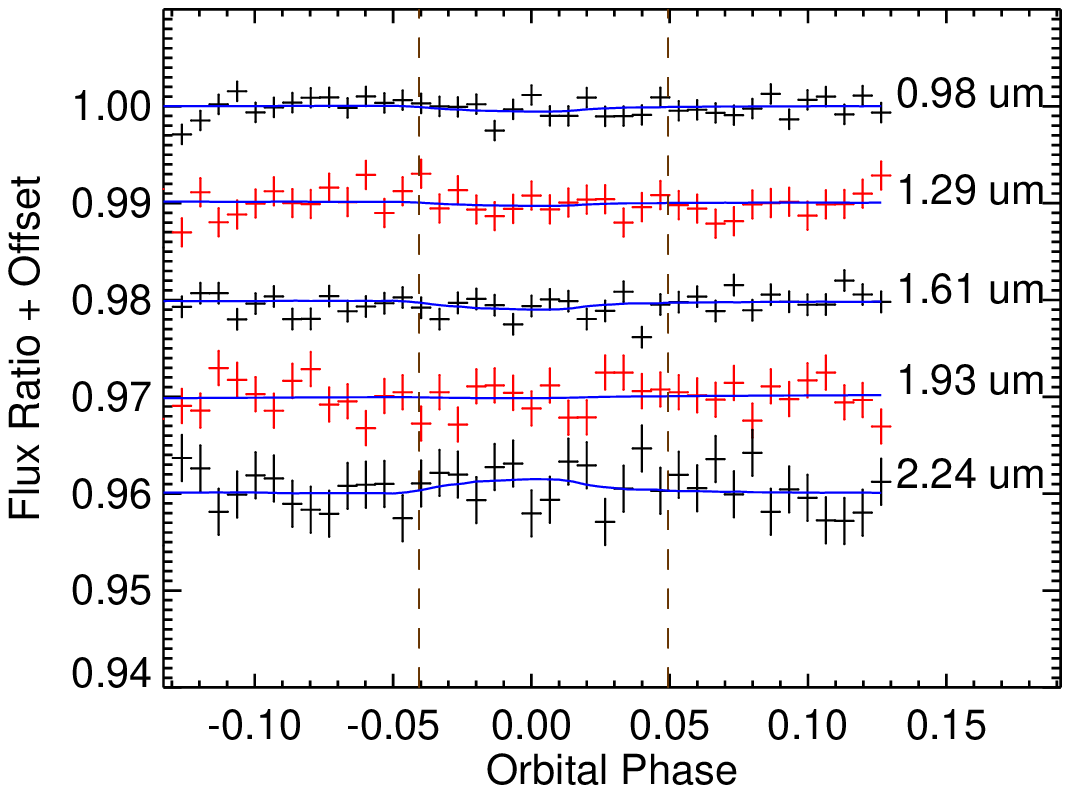}
\\
\includegraphics[width=0.48\textwidth]{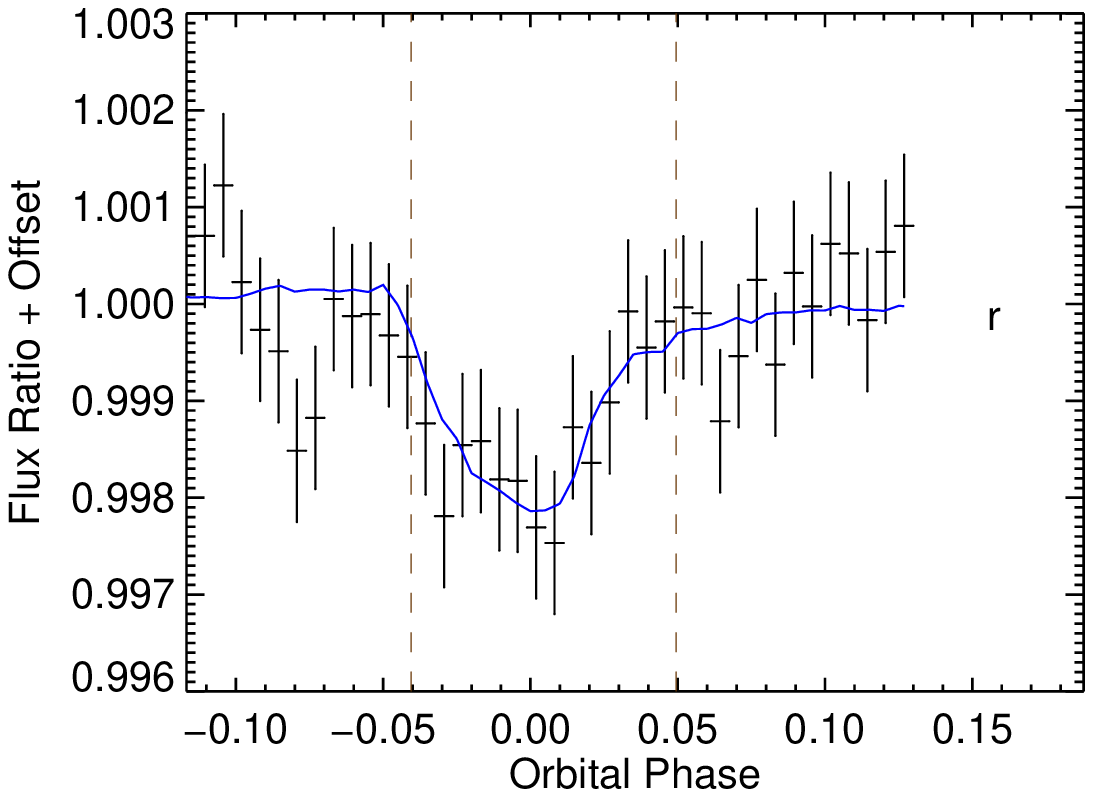}
\end{center}
\caption{{\it Top:} The time series for five equally spaced wavelength spectroscopic bins separated into 40 equally spaced time bins, 5.4 minutes each over the 5 nights with the largest MORIS transit depths. Each time series is divided by the broadband light curve to remove common-mode systematics. The average Kepler short cadence light curve (KSC) is used as a model (solid lines) with a free parameter for the scaling of the transit depth and two free parameters for the linear baseline, showing no statistically significant difference in transit depth from one wavelength to another in the best fits to the differential time series (blue lines). Red/black colors are alternated between wavelengths to distinguish overlapping time series. \newline {\it Bottom:} The MORIS $r'$ photometry ($\sim$ 0.56 to 0.69$\mu$m) for the 5 deeper transit depth nights is averaged together and also is fit with the KSC curve. The average transit depth for these nights when fit individually is 0.25\% $\pm$ 0.03\%, about half the average transit depth measured from the Kepler short cadence data (0.50\%).}\label{fig:timeSer5bin}
\end{figure}

\subsection{Light Curve Fitting}\label{sec:lcFit}

We averaged all available short cadence Kepler data (quarters 13 through 17 or 13 months) in order to construct a light curve model. We phased the light curve at an orbital period of P=0.6535538 days \citep{vanWerkhoven2014} and binned the data into time samples of 9.4 minutes (0.01 in orbital phase), which results in median errors of 0.005\%, well below the measured precision of SpeX or MORIS. There are still small residuals in the curve, visible out of transit in Figure \ref{fig:PhotIndNights} for September 3, 2014.

\begin{figure}[!ht]
\begin{center}
\includegraphics[width=0.5\textwidth]{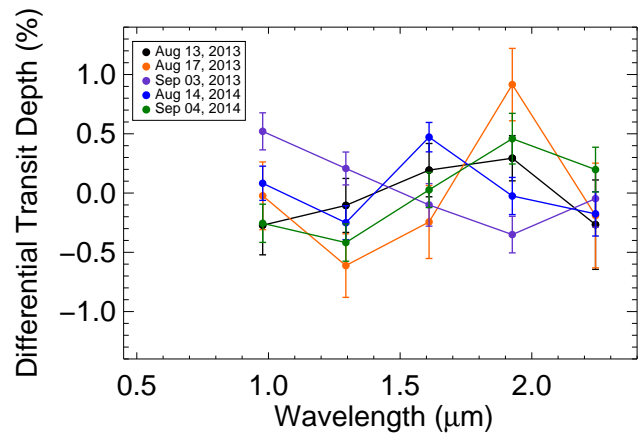}
\end{center}
\caption{Individual differential spectra are found from the SpeX data covering the 5 nights with the deepest transit depths as measured by the MORIS camera in the $r'$ band. The transit depth and errors are found by scaling the average Kepler Short Cadence (KSC) curve to best fit the data. The individual spectra broadly agree to within the errors.}\label{fig:IndDiffSpecFits}
\end{figure}

We fit each time series for each night individually using this average Kepler Short Cadence light curve (KSC), but scaled from its out-of-transit value by a free parameter. We bin the spectrum into five wavelength bins to improve the signal to noise for model fitting. We also include a linear baseline (fit simultaneously with the model as in \citet{schlawin2014}) to account for long term trends. The transit epoch is fixed at a constant-period ephemeris with mid-transit at 2454833.039 BJD, the minimum of the Kepler short-cadence average light curve. We also confirmed that the average MORIS light curve is consistent with this ephemeris within errors when the epoch is left as a free parameter. Error bars in the time series are calculated as the standard deviation of the flux out of transit. These are propagated to transit depth scaling with a numerical covariance matrix using the routine \texttt{mpfit} \citep{markwardt2009mpfit}. We down-selected the 5 nights with transit depths above 0.1\% as measured by the MORIS camera in $r'$ band to use for light curve analysis. Illustrative fits are shown in Figure \ref{fig:timeSer5bin} for the average time series of these nights both for the differential time series and the photometric light curve. The individual extracted spectra, shown in Figure \ref{fig:IndDiffSpecFits}, are within errors of each other and show no consistent trend with wavelength.

In order to calculate the inferred dust particle sizes from the differential spectra, it is necessary to assume a mean value. We adopt the mean value equal to the average MORIS $r'$ band transit depth for these 5 larger transit nights or \diffconst. This assumption is based on (1) the fact the optical ($0.42 \mu$m to 0.90$\mu$m) to infrared (2.15 $\mu$m) transit depth ratio is 1.02 $\pm$ 0.20 during two different transits \citep{croll2014} (2) the measured spectrum is shown to be flatter during shallow transit events \citep{bochinski2015evolving} and (3) our simultaneous optical (0.63$\mu$m) and H band (1.63 $\mu$m) measurement on August 18, 2014 are consistent to one another within the errors (see Figure \ref{fig:PhotIndNights}, middle right panel).

\begin{figure}[!ht]
\begin{center}
\includegraphics[width=0.5\textwidth]{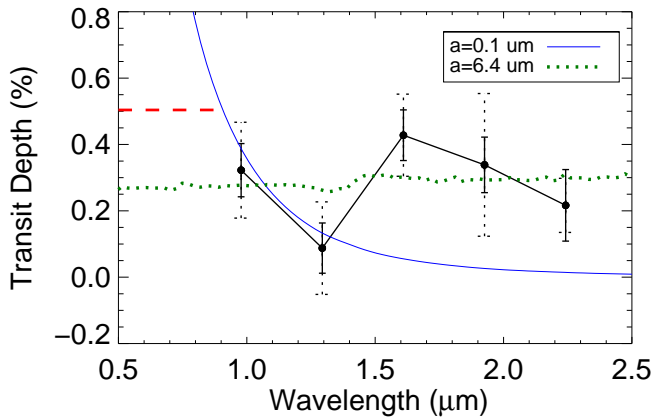}
\end{center}
\caption{The average differential spectrum of KIC 12557548b from the individual nights shown in Figure \ref{fig:IndDiffSpecFits}. The two error bars are the error propagation method (solid black error bars) and from the scatter in the individual nights (dotted black error bars). The red dashed horizontal line shows the average transit depth measured by the Kepler observatory in the optical. Our $r'$ photometry is far below the average Kepler value and the infrared spectrum is flat to within errors. Over-plotted are two example model spectra - one for a log-normal population of particles with a 0.1$\mu$m median radius and another for a larger 6.4$\mu$m median radius for Mg-rich pyroxene grains.}\label{fig:transmissionSpec}
\end{figure}

We combine the individual nights' spectra with a weighted average where the weights are proportional to the 1/$\sigma^2$ where $\sigma$ is the error in the transit depth. Next, we add the MORIS average $r'$ transit depth of \diffconst\ for the absolute offset. The resulting transmission spectrum is shown in Figure \ref{fig:transmissionSpec}, with two sets of error bars. The set of solid error bars is from the out-of-transit standard deviation propagated through to transit depth and the average. The dashed error bars are a more conservative estimate from the scatter in the data to calculate the standard deviation of the mean.

The average spectrum shown in Figure \ref{fig:transmissionSpec} favors large particle sizes, in contrast to our initial prediction that it would show strong wavelength dependence due to small particles. The forward scattering peak preceding transit is best fit by particle sizes of ${\sim0.1\mu\mathrm{m}}$ \citep{budaj12,brogi2012}, which have a pronounced drop in extinction from the optical to infrared wavelengths. Representative models are shown in Figure \ref{fig:transmissionSpec} for a log-normal distribution of spherical Mg-rich pyroxene grains. The model with a small population of grains represented by a fiducial $0.1 \mu$m median particle size (blue curve in Figure \ref{fig:transmissionSpec}) is disfavored by our data. For this model, the reduced chi-squared, $\bar{\chi^2}$, is 10.6. Instead, a population of larger population grains represented by a fiducial 6.4$\mu$m median radius grains (green dotted curve) better fits the data ($\bar{\chi^2}  = 2.6$). The models are described further in Section \ref{sec:psizes}.

\subsection{Inferred Particle Sizes}\label{sec:psizes}
We compare measured spectra to Mie scattering models for different compositions of dust grains. The calculated transmission spectrum does not have a steep downward slope from the optical to infrared as would be expected from small (0.1$\mu$m-sized) particles, shown in Figure \ref{fig:transmissionSpec}. To calculate the theoretical Mie spectra, we use the \texttt{IDL} Mie theory code \texttt{mie\_single.pro} \citep{grainger04} to calculate the extinction as a function of wavelength. We assume that the debris escaping from KIC 12557548 is optically thin and input the complex indices of refraction from \citet{dorschner95pyrox} for pyroxene and olivine compositions. We use \citet{ordal1998opticalconst} and \citet{koike1995corundum} for the complex indices of iron and corundum, respectively. For the particle size distribution, we consider a broad log-normal function with particle number proportional to $\exp{\left(-2 \left(\ln(a) - \ln(\mu)\right)^2 \right)}$, where $a$ is the particle radius and $\mu$ is the median of the distribution.

\begin{figure}[!ht]
\begin{center}
\includegraphics[width=0.47\textwidth]{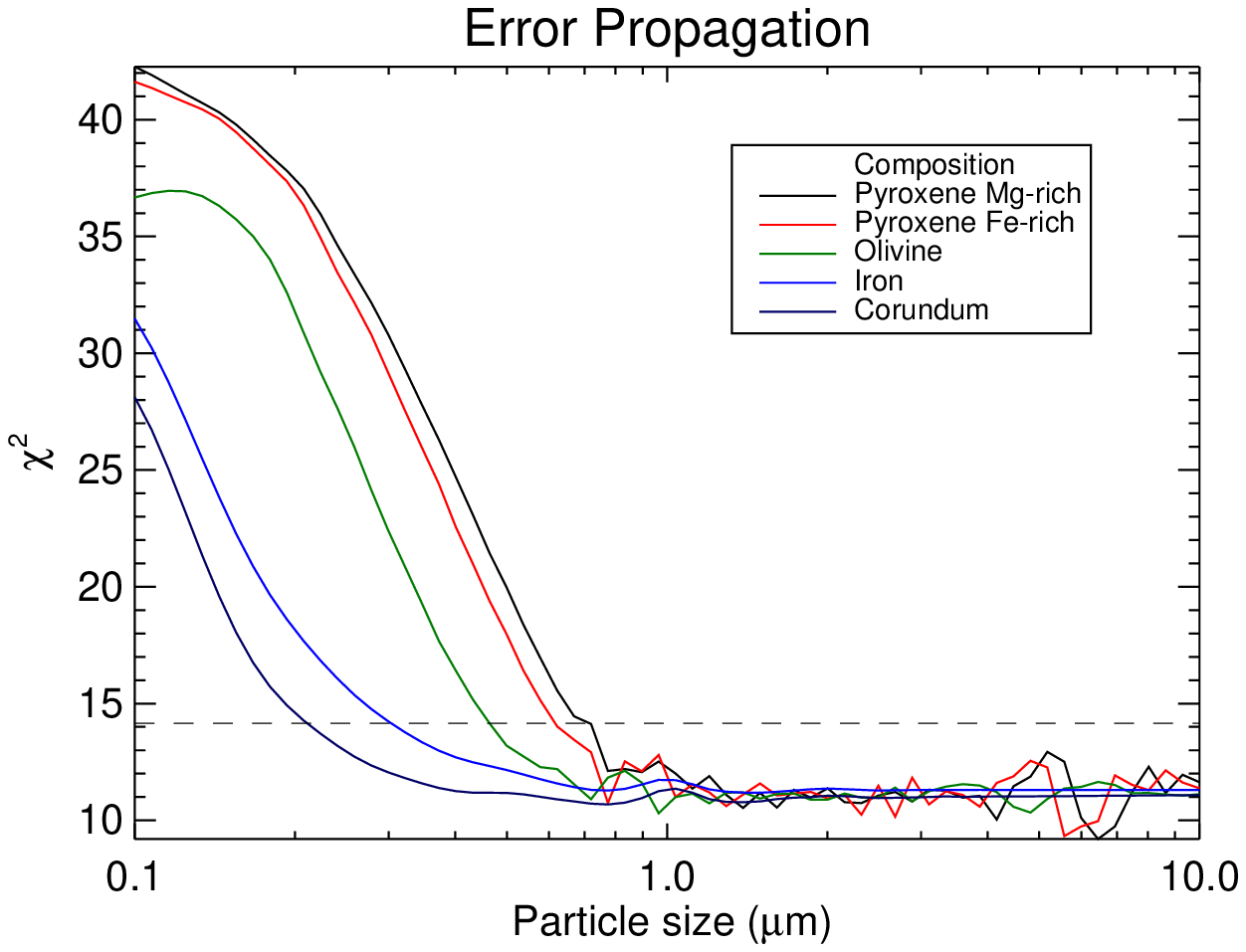}
\\
\includegraphics[width=0.47\textwidth]{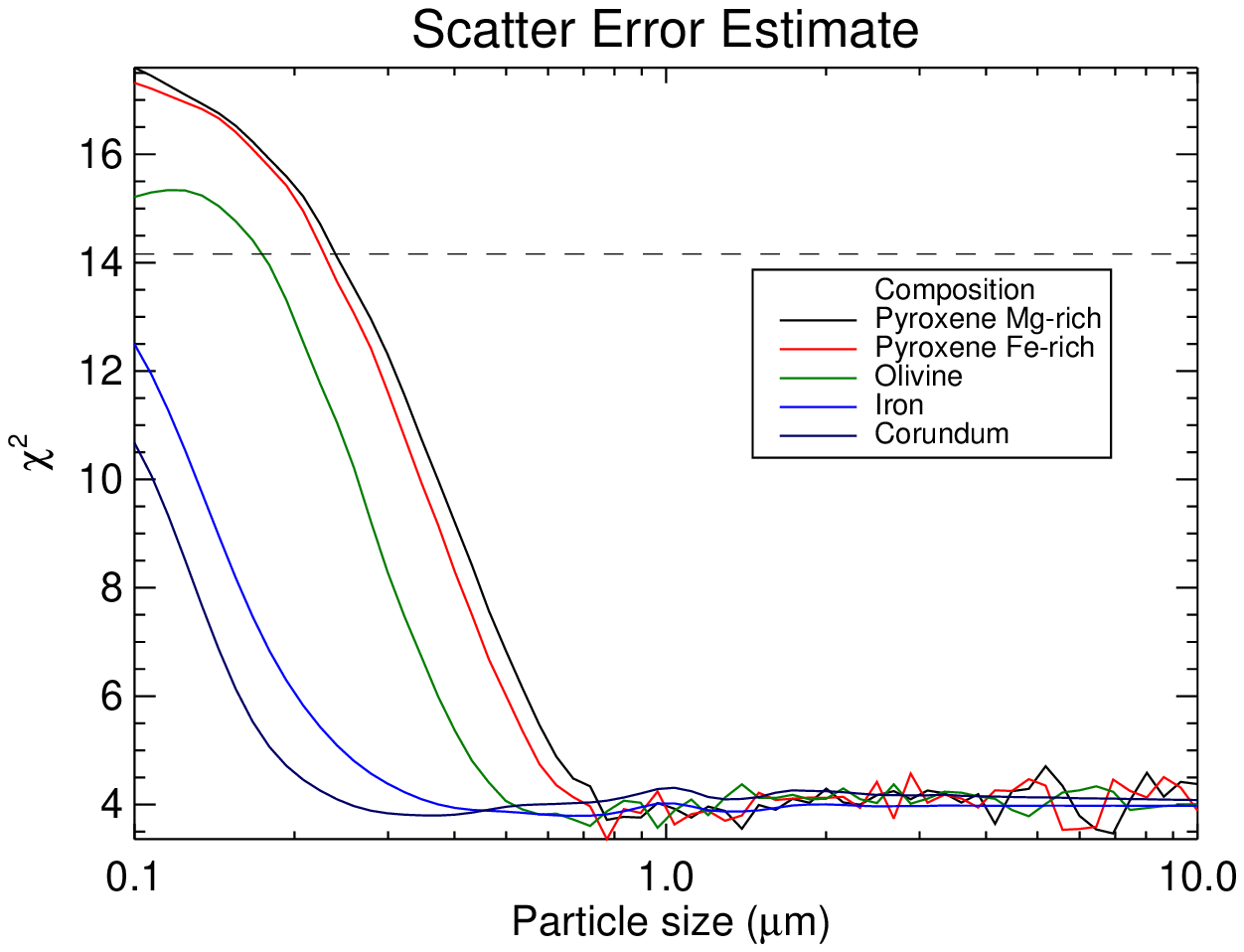}
\end{center}
\caption{$\chi$-Squared statistic as a function of median particle size for the measured differential spectrum shown in Figure \ref{fig:transmissionSpec} and a log-normal distribution of particle sizes. We consider three different compositions, but the inferred particle sizes are relatively insensitive to composition. A horizontal dashed lined shows the 0.2\% probability for a $\chi^2$ distribution with 4 degrees of freedom, which corresponds to the likelihood of a 3$\sigma$ event in a Normal distribution. The data favor large particle size distributions, but the significance varies on whether one adopts the propagated errors ({\it Top}) versus the scatter in the differential spectra ({\it Bottom}). }\label{fig:chisqPlot}
\end{figure}

In Figure \ref{fig:chisqPlot}, we show the $\chi^2$ statistic as a function of particle size for two different error estimates. For both error estimates and five different compositions: Mg-rich pyroxene (MgSiO$_3$), Fe-rich pyroxene (Mg$_{0.7}$Fe$_{0.3}$SiO$_6$) and olivine MgFeSiO$_4$ \citep{dorschner95pyrox}, pure iron and corundum (Al$_2$O$_3$) \citep{koike1995corundum}, the models with large dust particles better fit the data. For the errors propagated from the out-of-transit standard deviation in the time series, the lower limit on the particle sizes is 0.5 $\mu$m for the silicate grains and 0.2 $\mu$m for iron and corundum grains. The limits are for a Pearson's $\chi^2$ test probability of 0.2\% (equivalent of a 3 $\sigma$ event in a normal distribution). On the other hand, if the more conservative error estimate is taken from the scatter in the individual spectra from Figure \ref{fig:IndDiffSpecFits}, the same lower limit is 0.2 $\mu$m for the median of the lognormal particle size distribution for silicate compounds and unconstrained for iron and corundum. 

The large particle sizes calculated from our transmission spectrum and the shallow transit depths we observed for the 8 different nights spread over 2 years leave a few possible scenarios for KIC 12557548b's lifetime. If the particle size distribution remains unchanged between deep transits and shallow transits, then the large dust particles (also found by \citet{croll2014} during 2 average-depth transits) imply a short lifetime of the planet. If the particles are $\gtrsim 0.5\mu$m, then the mass loss rate of the system approaches 0.1M$_\oplus$/Gyr \citep{perez-becker}. For a maximum mass of 0.02M$_\oplus$, the lifetime of the planet $M/\dot{M}$ is less than 200 Myr, suggesting KIC 12557548b is in the late catastrophic phase of planet disintegration \citep{perez-becker}. On the other hand, the dust  size distribution may shift to smaller particle radii for deep transit depths as suggested by \citep{bochinski2015evolving}. In this case, the lifetime could be closer to 1 Gyr if it returns to $\sim 0.1 \mu$m particles for strong disintegration events. This would increase the likelihood of detecting disintegrating planets like KIC 12557548b, KOI 2700b and K2-22b. Similarly, if the disintegration activity has decreased significantly since since 2009-2013 when the Kepler Observatory monitored KIC 12557548, then the total mass ejected into the comet-like tail per orbit is less than about 1/3 the average value implied by the Kepler Quarters 1-17. This would increase the lifetime of the planet and mean that the Kepler Observatory happened on a outburst of disintegration that is less than the average activity over long ($\gtrsim$ 10yr) timescales.

\section{Conclusion}\label{sec:kic1255conclusion}
We observed the KIC 12557548 system for 8 transits using the SpeX spectrograph simultaneously with the MORIS imager in the $r'$ photometry band. Our observations took place after the Kepler observatory stopped monitoring KIC 12557548, so we relied on MORIS for optical photometry while taking spectroscopy from 0.8$\mu$m to 2.4$\mu$m with SpeX. The $r'$ photometric light curves have systematically smaller transit depths than the average value measured by the Kepler observatory. If we use the histogram from \citet{vanWerkhoven2014}, the probability of 8 independent transits having depths below 0.43\% is 0.2\%, indicating that the disintegration changed modes from when KIC 12557548 was observed by the Kepler observatory or that both in August/September of 2013 and 2014, our observations fell into particularly weak periods.

Our spectroscopic extractions using the SpeX instrument and differential spectroscopy show a flat spectrum across the wavelengths from 0.8$\mu$m to 2.4$\mu$m. The absolute time series were affected by systematics and had large variability in transit depth over the nights. This was unlike our previous observations of CoRoT-1b \citep{schlawin2014}, which achieved sub-0.1\% precision in transit depth across multiple wavelength channels. When we combine all of the differential spectra of KIC 12557548b over the nights with MORIS $r'$ transit depths greater than 0.1\%, we find that the transmission spectrum disfavors Mie scattering by small $\sim 0.1\mu$m olivine or pyroxene dust particles predicted by forward scattering models \citep{budaj12,brogi2012}. It is likely that the dust particles we observed are larger in size ($\gtrsim 0.5 \mu$m for silicates and $\gtrsim 0.2 \mu$m for iron and corundum) for our epochs of observation. A similar large particle size is found by \citet{croll2014}. Large dust particle sizes would result in short lifetimes of the planet, $\lesssim 10^7$yr \citep{perez-becker} if the dust particle size distribution we observed during shallow transits is the same as during deep transits.

Continued monitoring in the visible will ascertain if the unusually weak disintegration activity observed in the $r'$ band is an indication that the disintegration mechanism has changed or if it went through two weak periods during both August/September of 2013 and 2014. Additionally, the brighter recently discovered K2-22 system \citep{sanchis-ojedak2-22} presents another case to explore the transmission spectrum of a disintegrating planet. The K2-22 system's transit profile is more symmetric than KIC 12557548, indicating another geometry and potentially different particle size distributions for the escaping debris. Furthermore, long wavelength observations, such as with JWST will be useful for constraining the composition of the escaping debris and thus reveal the chemical makeup of these disintegrating bodies, a rare glimpse into the core of planets that have been peeled away layer by layer.

\section{Acknowledgements}

The authors wish to thank Bryce Croll and Saul Rappaport for their helpful suggested regarding observing, interpretation of the spectra, explanations of systematics, and wider range of compositions to consider. Thanks to James P. Lloyd and Dong Lai for their advice on the interpretation of the transits. Based on observations made from the Infrared Telescope Facility, which is operated by the University of Hawaii under contract NNH14CK55B with the National Aeronautics and Space Administration. M.Z. acknowledges support from NASA OSS grant NNX14AD22G and the Center for Exoplanets and Habitable Worlds at the Pennsylvania State University. The authors wish to recognize and acknowledge the very significant cultural role and reverence that the summit of Mauna Kea has always had within the indigenous Hawaiian community.  We are most fortunate to have the opportunity to conduct observations from this mountain. 

\appendix

\begin{figure*}[!ht]
\begin{center}
\includegraphics[width=0.7\textwidth]{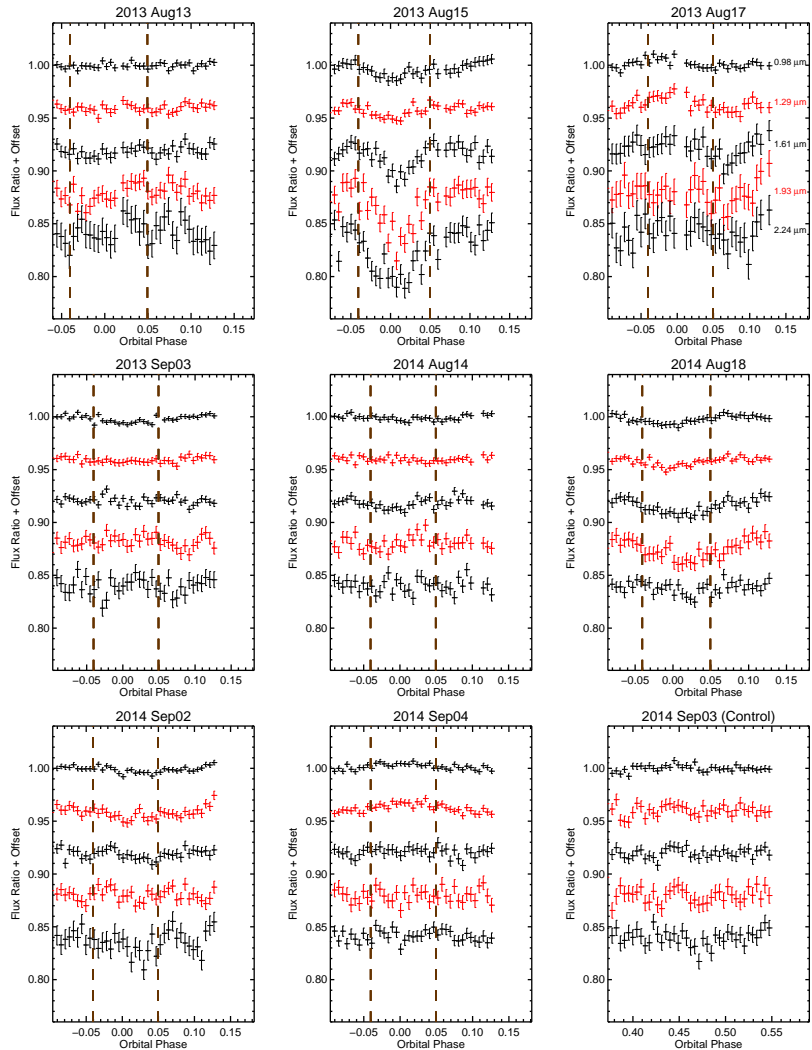}
\end{center}
\caption{SpeX time series binned into 5 wavelength bins for all nights, showing large systematic effects in the absolute time series and a likely false transit on August 15, 2013. Ingress and egress are marked by dashed horizontal brown lines. For analysis of the spectrum further on, these systematics are mitigated by dividing all light curves by the broadband average light curve for that night and by combining together data from multiple nights. Some gaps are visible in the light curves due to telescope guiding errors (August 13, 15, 17, 2013 and August 18, 2014) and setting an incorrect number of exposures initially (August 14, 2014). Eight of the observations took place during transit while September 3, 2014 was used as a control for understanding systematics.}\label{fig:dynamicSpecInd}
\end{figure*}

\section{Absolute SpeX Spectra}\label{sec:altAvgMeth}
\begin{figure}[!ht]
\begin{center}
\includegraphics[width=0.47\textwidth]{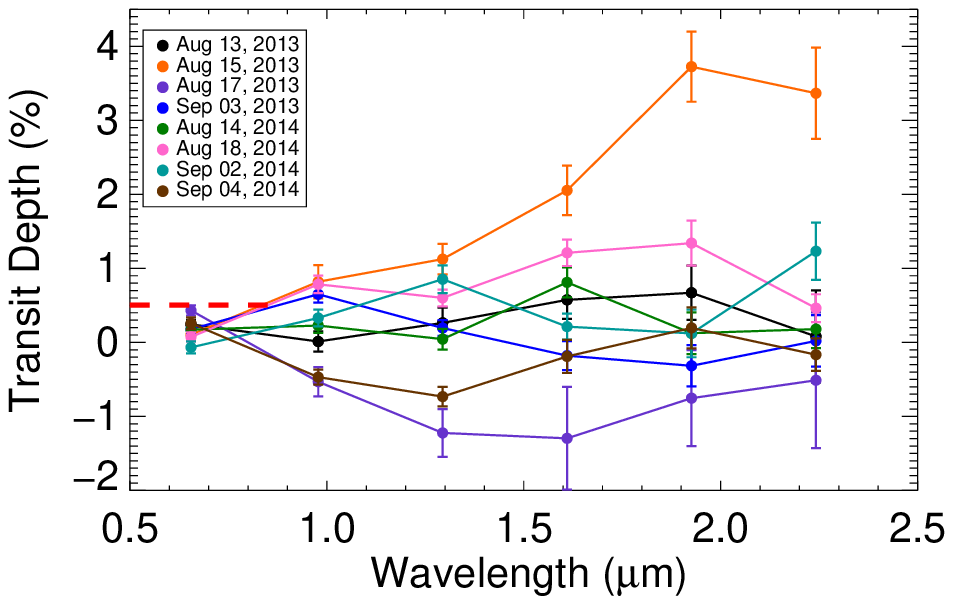}
\includegraphics[width=0.47\textwidth]{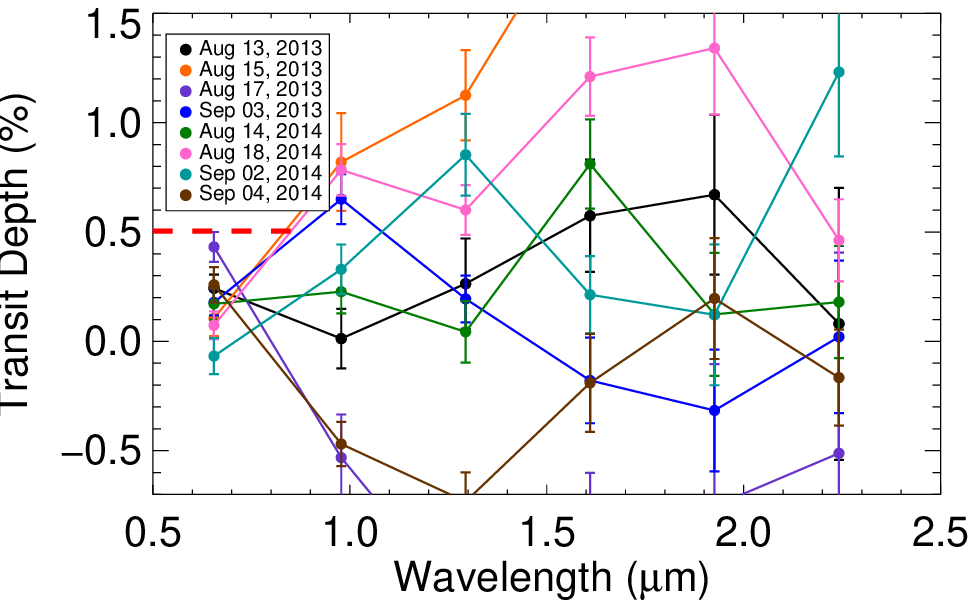}
\end{center}
\caption{{\it Left} Transmission spectra for all 8 nights individually, showing large variations far above the error bars estimated from out-of-transit errors propagated to transit depth. In all plots, the horizontal dashed line is the average transit depth measured by the Kepler observatory. {\it Right} Same plot, but zooming in.}\label{fig:spectrumIndNight}
\end{figure}

The individual transits of KIC 12557548b show dramatic night-to-night anomalies and variations in the SpeX infrared data, as visible in Figures \ref{fig:dynamicSpecInd} and \ref{fig:spectrumIndNight}. The system is intrinsically variable, as measured in high precision with the Kepler Observatory, although we expected that the variations among the 8 observed transits to be small because the MORIS photometry shows roughly 0.25\% or below transit depths for all nights. If we assume that the near-infrared optical depth is proportional to the optical depth at $\sim0.6 \mu$m (as found tentatively with the Palomar $H$ band data and by \citet{croll2014}), then the variability observed in the SpeX spectrograph from 0.8$\mu$m to 2.4$\mu$m is likely due to systematic errors. Still, we consider the possibility that some of these variations are astrophysical. In this section, we fit the time series in the absolute sense instead of differentially (as done in Sections \ref{sec:transm} and \ref{sec:psizes}.)

The nights of August 15, 2013 and August 17, 2013 show deviation in the relative flux of the background near the two stars within the SpeX instrument corresponding to the time of the transit. These background fluctuations likely created anomalous bumps and troughs in the SpeX light curve, producing a deep positive transit and negative transit respectively. Given the background fluctuations, we only include the 6 remaining nights for absolute transmission spectrum analysis.

\begin{figure}[!ht]
\begin{center}
\includegraphics[width=0.47\textwidth]{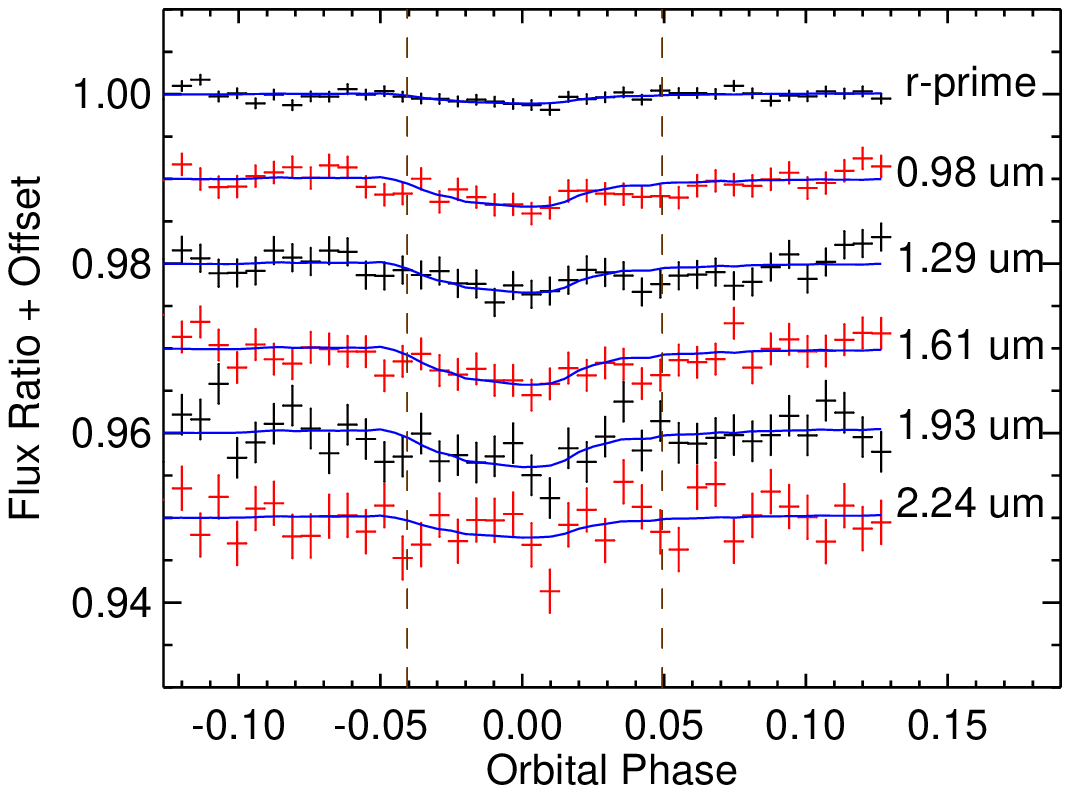}
\includegraphics[width=0.47\textwidth]{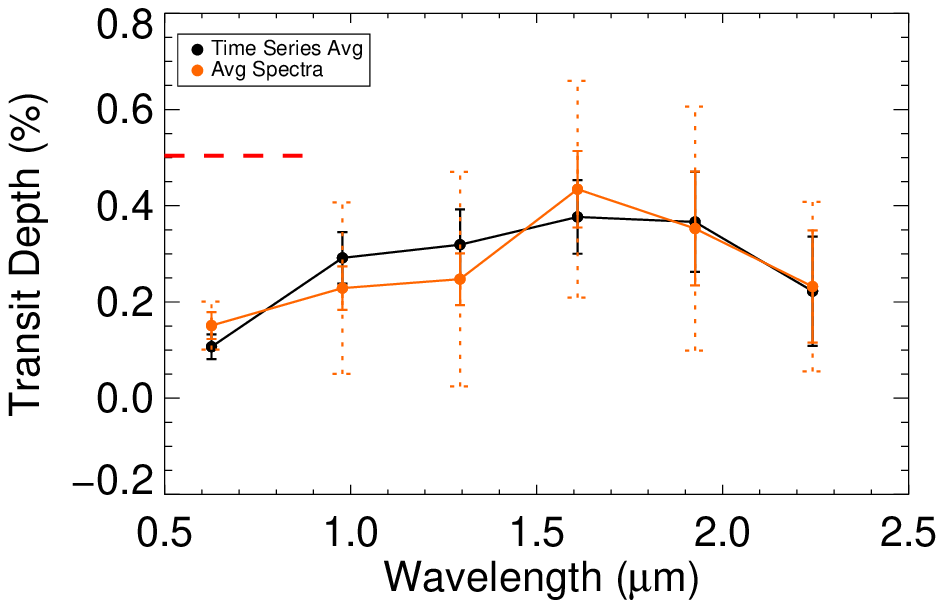}
\end{center}
\caption{{\it Left} The average time series for all 6 of the 8 nights fit with the KSC curve by scaling the transit depth as a free parameter. We exclude the nights of August 15, 2013 and August 17, 2013, which were affected by background fluctuations. {\it Right} The fit to the average time series and alternatively the average of individual nights' spectra, show a rising spectrum from the optical to infrared, likely due to common-mode systematic errors.}\label{fig:medianandTimSerAvg}
\end{figure}

We adopt several methods of averaging both the individual spectra shown in Figure \ref{fig:spectrumIndNight}. When combining spectra obtained on different nights, we used weighted averaging with the weights proportional to $1/\sigma^2$, where $\sigma$ is the error in transit depth propagated from the out-of-transit standard deviation. When estimating the error of the average spectra, we also calculate the standard deviation of the mean using the scatter in the data as a more conservative error estimate. Alternatively, one can combine the time series of multiple nights (shown in Figure \ref{fig:medianandTimSerAvg} {\it Left}) and fit this combined time series. In the average time series method, errors are calculated from the standard deviation of the out-of-transit flux. The time series method results in a spectrum that rises from the optical $r'$ band to the infrared, shown in Figure \ref{fig:medianandTimSerAvg} ({\it Right}).

\subsection{Alternate Chi-Squared Particle Size Constraints}\label{sec:altPSizeConstr}

\begin{figure}[!ht]
\begin{center}
\includegraphics[width=0.47\textwidth]{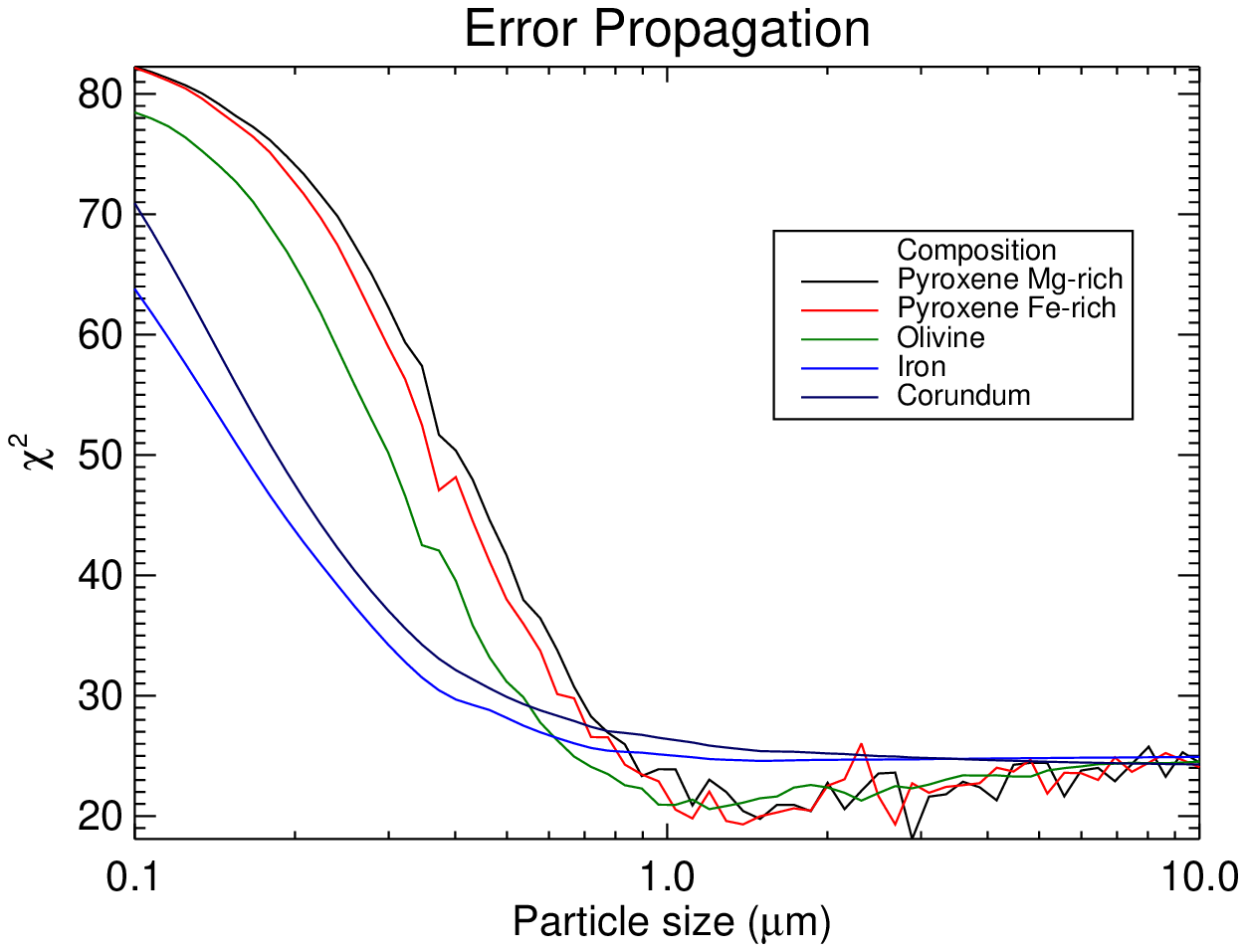}
\includegraphics[width=0.47\textwidth]{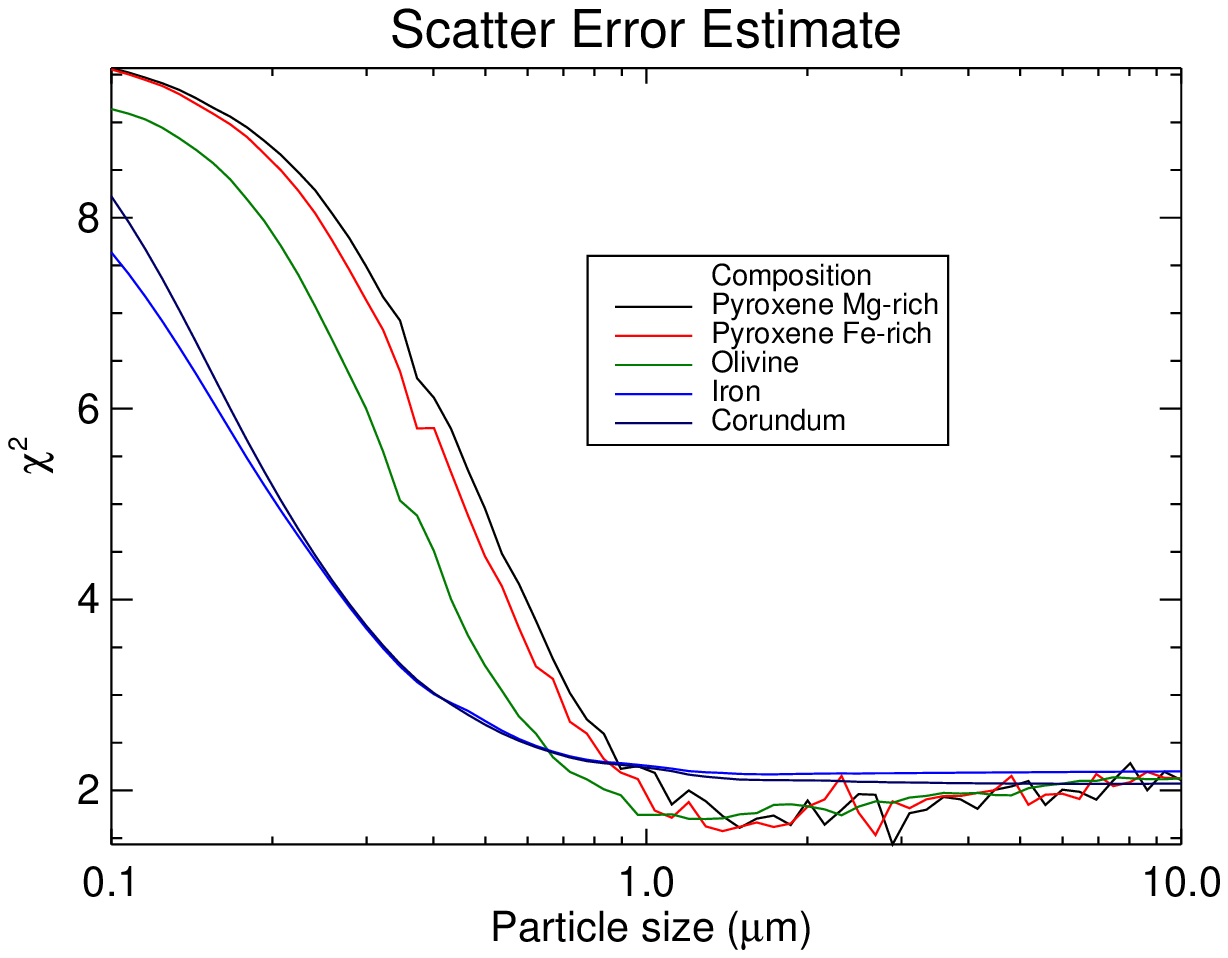}
\end{center}
\caption{{\it Left} Chi-Squared statistic as a function of particle size for the measured 6 night time series average in Figure \ref{fig:medianandTimSerAvg} and a log-normal distribution of particle size. The data favor large particle size models, but none can explain the the rise in transit depth with wavelength. {\it Right} Same statistic, but this time using a average of the 6 spectra and the scatter amongst the nights as a more conservative error estimate. The particle size is only weakly constrained.}\label{fig:chisqPlotAlt}
\end{figure}

For the average absolute spectra shown in Figure \ref{fig:medianandTimSerAvg} we fit the data to Mie scattering models with log-normal size distributions of spherical pyroxene, olivine, iron and corundum grains, as done in Section \ref{sec:lcFit}. We show the constraints on particle sizes in Figure \ref{fig:chisqPlotAlt} for the 6 night time series average and for the average of the 6 nights with errors estimated from the scatter in the nights. In all cases, as with the differential spectra, the data favor large dust particle sizes, but the significance of the constraints changes depending on the method. Using the scatter in the spectra (more conservative case) gives the maximum error bars since it assumes no intrinsic astrophysical variability to the system. Continued monitoring of KIC 12575548b in both the infrared and optical will be invaluable in constraining the particle size and degree of variability.

\subsection{Transit Injection}\label{sec:TransitInject}
\begin{figure}[!ht]
\begin{center}
\includegraphics[width=0.47\textwidth]{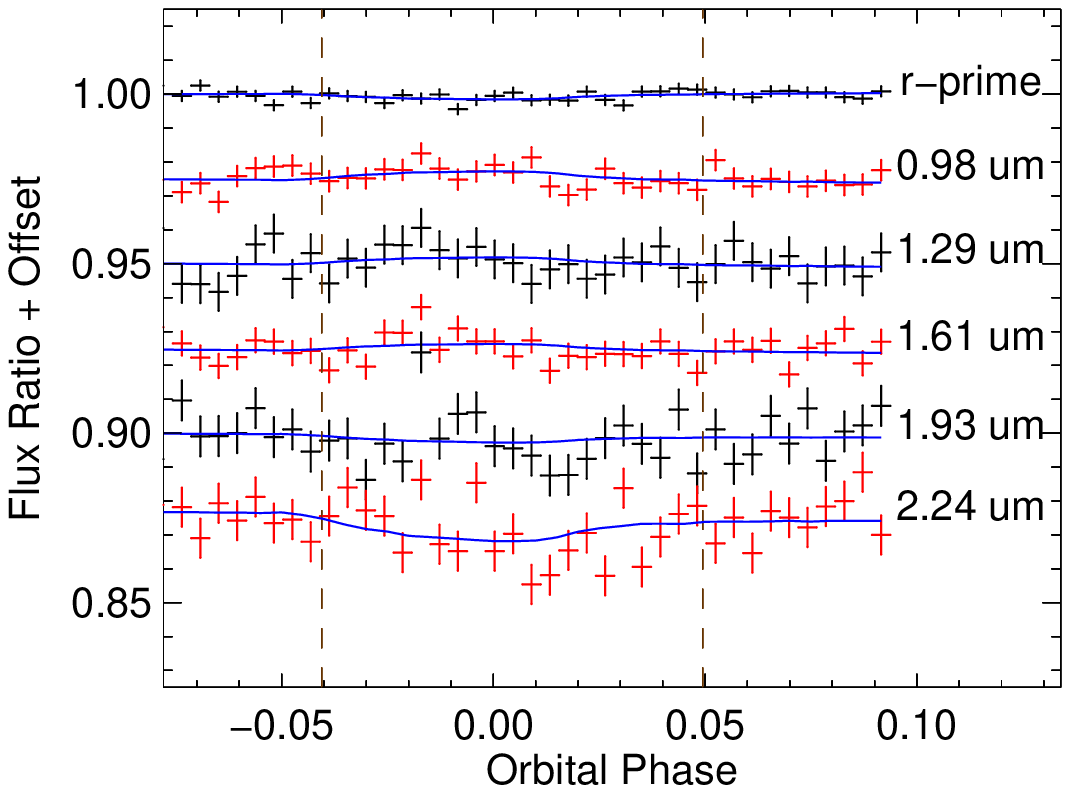}
\includegraphics[width=0.47\textwidth]{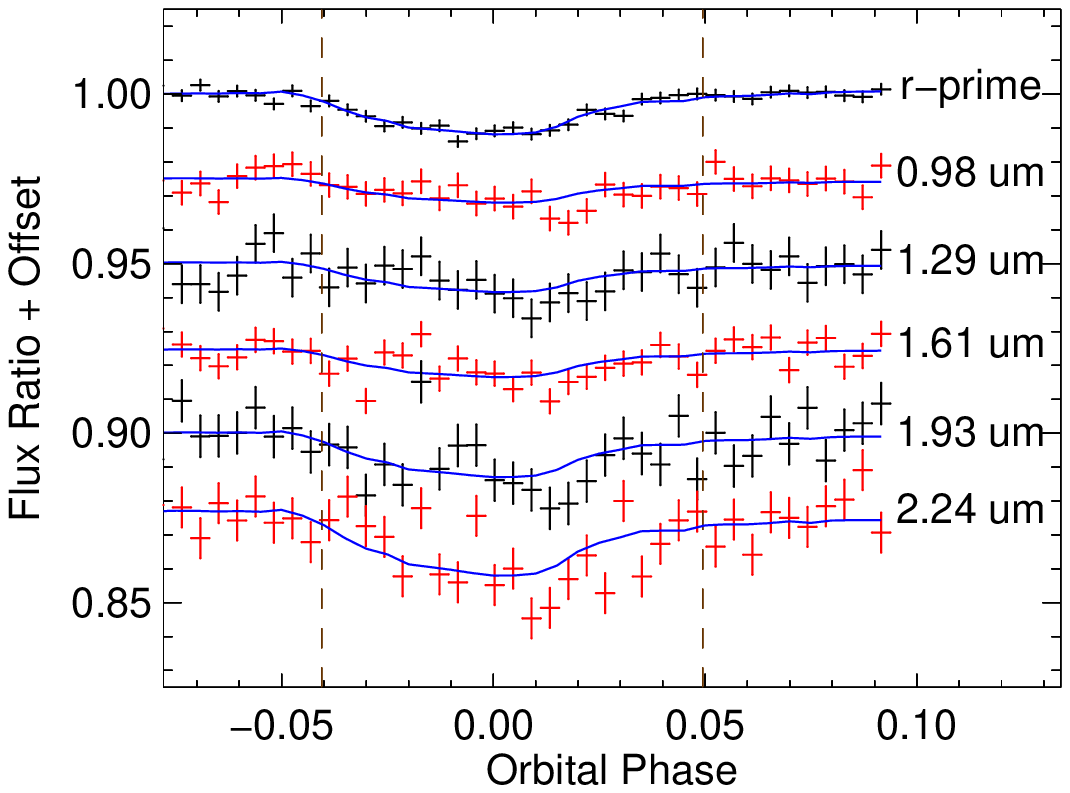}
\end{center}
\caption{{\it Left} The control night of September 3, 2014 where the orbital phase has been shifted to estimate the errors in the fitting process. The control night is fit with the KSC light curve scaled to best fit the transit depth and does give deviations from the expected value of zero. {\it Right} If we artificially inject a 1\% wavelength independent transit into the control night (a typical deep Kepler transit depth), it is recovered, though the fitted transit depths deviate from a flat spectrum.}\label{fig:transitInjectRec}
\end{figure}

As mentioned in Section \ref{sec:lcFit}, the control night of September 3, 2014 is fit as if it occurred during a transit to determine if there are systematics. The fitted transmission spectrum for this control night is not zero within errors, but the systematics are smaller than the large deviations occurring on August 15, 2013 and August 17, 2013. The time series fits are shown in Figure \ref{fig:transitInjectRec} left, where we take the control night and shift the orbital phase to fit a KSC model as if it were real data during a transit. We perform another test of significance by injecting a transit into our control night of data, September 3, 2014 at a reference epoch of BJD= 2456903.8. This control night shows deviations from zero that are not expected for Gaussian identically distributed independent errors, as listed in Table \ref{tab:controlSpec}. Despite these variations, we demonstrate that at 1\% deep transit would be recovered by the SpeX spectrograph and MORIS $r'$ imager, shown in Figure \ref{fig:transitInjectRec} right.

\section{Bootstrap Error Estimates}\label{sec:bootstrapErr}
\begin{figure}[!ht]
\begin{center}
\includegraphics[width=0.47\textwidth]{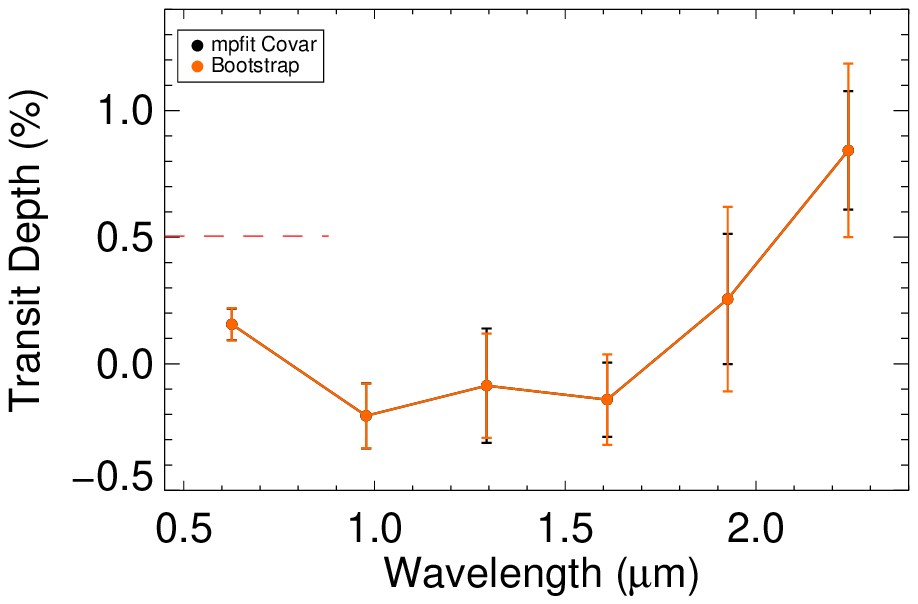}
\includegraphics[width=0.47\textwidth]{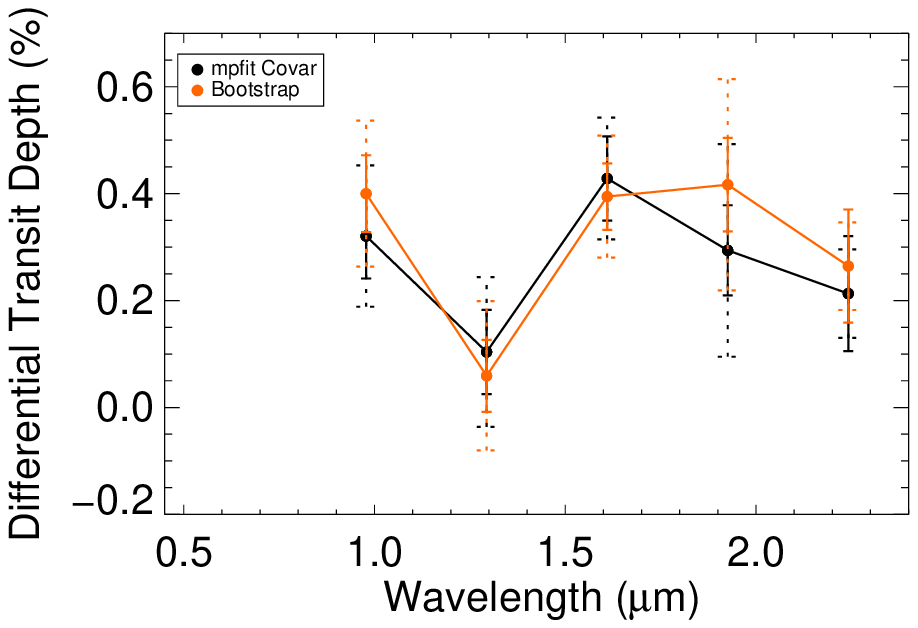}
\end{center}
\caption{{\it Left} The absolute spectrum of the control night of September 3, 2014 with two estimates for error bars. Bootstrap resampling gives comparable errors (though larger at long wavelengths) than propagating the out-of-transit standard deviation error estimates in the time series to the transit depth with a covariance matrix within \texttt{mpfit} {\it Right}. The differential spectrum combined over all nights, as in Figure \ref{fig:transmissionSpec}. The same two methods are used to estimate the uncertainty in transit depth for individual nights' differential transmission spectra. We combine the nights with a weighted average and show that the final differential transmission spectrum is relatively unchanged, favoring large particles ($\gtrsim 0.5 \mu$m) in the escaping winds from KIC 12557548b. As in the other plots, the dashed error bars are for errors estimated from the scatter in the individual nights' spectra.}\label{fig:bootstrapMpfitCov}
\end{figure}

Another way to estimate errors in the transit depth from the light curve is by bootstrapping \citep[e.g.][]{freedman1981bootstrap}. We resample the data with replacement so that the new time series is as long as the original. We find the best fit and repeat for 500 different re-samples of data of each time series. We then calculate a standard deviation of the fitted transit depths to find an uncertainty in the best-fit transit depth. The errors are close to the method described in Section \ref{sec:lcFit} using a covariance matrix from \texttt{mpfit}, as visible for an example of the control night fit in Figure \ref{fig:bootstrapMpfitCov} {\it(left)}, though they are larger for long wavelengths. We also fit all differential transit depths with bootstrap errors and calculate the weighted average of the 5 nights with the deepest MORIS $r'$ transits as done in Section \ref{sec:transm}. This weighted average differs slightly from the method using the \texttt{mpfit} covariance matrix by as much as 1.2$\sigma$, but the final result of the flat transmission spectrum (which favors large dust particle sizes $\gtrsim 0.5 \mu$m) remains unchanged, visible in Figure \ref{fig:bootstrapMpfitCov} Right.

\section{Demonstration of MORIS Photometry on a Faint Source}
\begin{figure}[!ht]
\begin{center}
\includegraphics[width=0.5\textwidth]{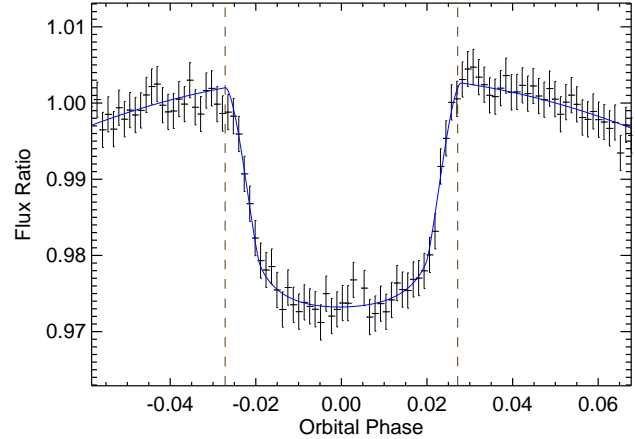}
\end{center}
\caption{Time series for CoRoT-2 (K=10.3) using a K=13.4 reference star to demonstrate MORIS photometric precision on a faint source. A quadratic baseline is used to model the low frequency modulation of the light curve. The best-fit transit depth has a 0.09\% precision, which is more than sufficient to detect average transits of KIC 12557548b at 0.5\%.}\label{fig:corot2fstarSeries1}
\end{figure}

KIC 12557548 and its reference star are fainter than previous high precision photometry and spectrophotometry used in \citet{schlawin2014}. To explore the precision that can be achieved with the MORIS camera photometry for a similarly faint system, we examine data on the hot Jupiter CoRoT-2b from UT July 4, 2012. For the test, we take the CoRoT-2 system (K=10.3) and divide it by a reference star (2MASS 19270544+0123119, K=13.436). The USNO A1 R-band fluxes are 15.5 both for 2MASS 19270544+0123119 and KIC 1255's reference star 2MASS J19234770+5130175, so the brightness-dependent systematics should be very similar with respect to the reference star. It should be noted that CoRoT-2 is significantly brighter than KIC 12557548 so it is not a perfect analog with respect to the target star. Figure \ref{fig:corot2fstarSeries1} shows the time series around transit. We find a transit depth of $(R_p/R_*)^2$ of 2.61\% $\pm$ 0.09\% consistent within errors of the CoRoT measurement of 2.75\% $\pm$ 0.01\% \citep{gillon2010corot2}. Our best-fit included a quadratic baseline, free quadratic limb darkening parameters, a free $Rp/R_*$ and fixed $a/R_*$ and impact parameter from the CoRoT-based values. The filter used was an LPR-600 light pollution reduction filter which is a broadband filter with reduced transmission at 0.6$\mu$m.

\bibliography{master_biblio}

\end{document}